\documentclass[11pt,a4paper,oneside]{article}
\usepackage{lineno,hyperref}

\usepackage{graphicx}              % to include figures
\usepackage{epsfig}
\usepackage{amsmath}               % great math stuff
\usepackage{amssymb}
\usepackage{epstopdf}
\usepackage{verbatim}
\usepackage{mathptmx}
\usepackage{mathalpha}
\usepackage{mathtools}
\usepackage[scale=0.80]{geometry}
\usepackage{graphicx, times}
\usepackage{amsfonts}              % for blackboard bold, etc
\usepackage{amsthm}                % better theorem environments\right 
\usepackage{multicol}
\usepackage{algorithm}
\usepackage{algorithmic}
\usepackage{varwidth}
\usepackage{parskip}
\usepackage{hyperref}
\usepackage{rotating}
\usepackage[numbers,sort&compress]{natbib}
\usepackage{multirow}
\usepackage{pdflscape}
\usepackage[numbers]{natbib}
\usepackage{caption}
\usepackage{xcolor}
\usepackage{color}
\usepackage{comment}
\usepackage{subcaption}
\newcommand*{\rom}[1]{\expandafter\@\romannumeral #1}
\newcommand{\bea}{\begin{eqnarray}}
	\newcommand{\eea}{\end{eqnarray}}
\newcommand{\bee}{\begin{eqnarray*}}
	\newcommand{\eee}{\end{eqnarray*}}

\begin{document}
\author{Khomesh R. Patle$^{}$\footnote{khomeshpatle5@gmail.com}, G. P. Singh$^{}$\footnote{gpsingh@mth.vnit.ac.in}
\vspace{.3cm}\\
${}^{}$ Department of Mathematics,\\ Visvesvaraya National Institute of Technology, Nagpur, 440010, Maharashtra, India.
\vspace{.3cm}
\date{}}
\title{Probing cosmic dynamics in $f(T)$ teleparallel gravity: Constraints from logarithmic and log-periodic deceleration ansatzes}
\maketitle
\begin{abstract} \noindent
In this study, we probe the cosmological evolution of the universe within the framework of modified teleparallel gravity by considering a power-law form of the function $f(T)=\alpha(-T)^{n}$. To characterize the expansion dynamics, we employ logarithmic and log-periodic parametrizations of the deceleration parameter. These specific parametrizations provide a flexible and well-structured description of the cosmic expansion history across different cosmological epochs. The corresponding Hubble parameter is obtained as a function of redshift, facilitating a systematic investigation of the background dynamics. The model parameters are constrained using cosmic chronometer (CC) and joint (CC+Pantheon) datasets through a Bayesian analysis based on the $\chi^{2}$-minimization approach. The evolution of key cosmological quantities, including the deceleration parameter, energy density, pressure, equation of state parameter and energy conditions, is examined in detail. The geometrical diagnostics indicate a clear departure from the standard cosmological constant behavior, pointing toward a dynamically evolving dark energy scenario. Further, the proposed models remain thermodynamically consistent and the estimated age of the universe is found to be compatible with observational constraints, thereby reinforcing the robustness and viability of the framework.
\end{abstract}
\section{Introduction}\label{sec:1}
Observational investigations from several independent cosmological probes~\cite{1998AJ....116.1009R,1999ApJ...517..565P,2020A&A...641A...6P} strongly support the view that the universe is presently evolving through an accelerated expansion phase. The discovery of this accelerated behavior has led to the emergence of the dark energy (DE) concept, which is regarded as the principal source responsible for driving the late-time dynamics of the universe. Although DE dominates the present cosmic energy content, its intrinsic physical interpretation remains uncertain and continues to be one of the central unresolved problems in cosmology. Observational studies further indicate that dark energy together with dark matter, constitutes nearly $95$–$96\%$ of the total energy density of the universe~\cite{weinberg1989cosmological}. Several theoretical mechanisms have been proposed to explain the origin of DE, among which the cosmological constant ($\Lambda$) stands out as the simplest and most phenomenologically successful description. The standard $\Lambda$CDM cosmological scenario agrees remarkably well with a large collection of observational measurements; however, it continues to suffer from serious conceptual issues, particularly the fine-tuning and cosmic coincidence problems~\cite{di2021realm,carroll2001cosmological,padmanabhan2003cosmological,copeland2006dynamics}. The existence of these theoretical shortcomings has stimulated the search for alternative cosmological frameworks capable of explaining the accelerated expansion history of the universe. In this direction, extensive studies have been devoted both to models containing unconventional matter sources and to theories based on modifications of Einstein’s General Relativity (GR). Among these alternatives, modified gravity approaches have received considerable attention because they can provide a geometric explanation for late-time cosmic acceleration without the need for an additional dark energy component. Consequently, a wide variety of modified gravity models has been constructed and explored in cosmological studies~\cite{buchdahl1970non,harko2011f,nojiri2011unified,jimenez2018coincident,nojiri2017modified,bamba2010finite,elizalde2010lambdacdm,harko2010f,capozziello2019extended,capozziello2023role,kotambkar2017anisotropic,lalke2023late,hulke2020variable,singh2025observational,singh2024conservative,chaudhary2025extracting,goswami2024flrw,shukla2025multi,escamilla2024exploring,patle2026dynamical}.
\vspace{0.2cm}\\
Among the various modified gravity frameworks proposed over the past few years, $f(T)$ teleparallel gravity~\cite{Bengochea,cai2016f} has been recognized as one of the most widely explored and promising alternatives. This framework generalizes the teleparallel equivalent of GR (TEGR) by extending the gravitational action from the torsion scalar $T$ to an arbitrary function $f(T)$, thereby constructing a broader class of torsion-based gravitational models. In this respect, it shares a formal resemblance with $f(R)$ gravity, in which the Einstein-Hilbert action is generalized by promoting the Ricci scalar $R$ to a function $f(R)$. Despite this formal similarity, the two frameworks are fundamentally different in their geometric foundations. GR is formulated using the torsion-free Levi-Civita connection, in which gravitational effects are described as a consequence of spacetime curvature. In contrast, teleparallel gravity is formulated using the curvature-free Weitzenböck connection, in which gravitational interactions are described entirely in terms of torsion rather than curvature. At the level of TEGR, this torsion-based description yields dynamics that are equivalent to those of GR. Nevertheless, when generalized to $f(T)$ gravity, the resulting field equations deviate from GR, giving rise to rich and nontrivial cosmological dynamics. Owing to its relatively simple mathematical structure and its ability to account for the late-time accelerated expansion of the universe without the need for an explicit DE term, $f(T)$ gravity has received significant interest in modern cosmological studies.
\vspace{0.2cm}\\
A substantial body of work has been devoted to examining the cosmological viability of $f(T)$ gravity from multiple viewpoints. These studies encompass analyses of dynamical cosmological models that characterize the evolution of the universe~\cite{paliathanasis2016cosmological}, together with examinations of thermodynamic properties in the context of teleparallel gravity~\cite{salako2013lambdacdm}. Model-independent approaches such as cosmography, have been widely employed to reconstruct the expansion history of the universe~\cite{capozziello2011cosmography}, while the validity of classical energy conditions has been analyzed to evaluate the physical consistency of multiple $f(T)$ gravity formulations~\cite{liu2012energy}. In addition, alternative cosmological pictures of the early universe, such as matter-bounce models, have been formulated within this framework~\cite{cai2011matter}. Evidence from observational investigations has further strengthened the motivation for studying $f(T)$ gravity. As an illustration, Zhadyranova et al.~\cite{zhadyranova2024exploring} investigated the late-time accelerated expansion of the universe in the context of a linear $f(T)$ model and constrained the corresponding model parameters using observational datasets. A comprehensive review of the theoretical foundations and cosmological consequences of $f(T)$ gravity is presented in Ref.~\cite{cai2016f}. In addition, Bamba et al.~\cite{bamba2011equation} explored the evolution of the DE EoS parameter ($\omega_{DE}$) for exponential, logarithmic and hybrid functional forms of $f(T)$. Paliathanasis et al.~\cite{paliathanasis2014new} also explored the Noether symmetries of the theory in detail, revealing important information regarding its conserved quantities and integrability structure. Moreover, Capozziello et al.~\cite{capozziello2017model} established a model-independent computational approach to solve the modified cosmological field equations in teleparallel gravity. Taken together, these investigations, along with many other studies available in the literature~\cite{shekh2025cosmographical,duchaniya2024attractor,maurya2023anisotropic,maurya2022accelerating,bamba2016bounce,kavya2024can,bhar2024anisotropic,nunes2016new,chaudhary2024constraints,duchaniya2022dynamical,chakraborty2023classical,maurya2024role,dixit2021probe,patle2026revisiting,das2023study,ren2022gaussian}, demonstrate the rich theoretical structure and strong observational viability of $f(T)$ gravity, highlighting its role as a well-founded and promising approach for addressing key challenges in contemporary cosmology.
\vspace{0.2cm}\\
The parameter $q(z)$, commonly known as the deceleration parameter, provides useful information about the evolutionary behavior of the cosmic expansion, as it directly distinguishes between decelerating $(q > 0)$ and accelerating $(q < 0)$ phases. Its evolution provides crucial insights into the progression from an early matter-dominated decelerated expansion to the current epoch of accelerated expansion, as strongly supported by various cosmological observations. Accurately capturing this transition is essential for understanding both the formation of large-scale structures and the physical origin of the accelerated expansion of the universe. To investigate the dynamical evolutionary pattern of the universe's expansion, numerous parametrizations of the deceleration parameter have been proposed in the literature. Such parametrizations range from simple linear representations like $q(z)=q_0+q_1 z$ to more generalized formulations, including CPL-type parametrizations as well as models involving rational and inverse redshift terms~\cite{riess2004type,gong2006observational,arora2024diagnostic,sanchez2011tracing}. Although these parametrizations have been widely employed, many of them are either too restrictive or applicable only over limited redshift ranges, thereby limiting their ability to fully describe the expansion history of the universe~\cite{gadbail2022parametrization,koussour2023modeling,shekh2024late,al2016divergence,xu2008constraints}. Motivated by these considerations, in the present work we adopt a logarithmic form of the deceleration parameter expressed as $q(z)=q_0+q_1\log(1+z)$~\cite{sofuouglu2026scalar}, which introduces a smooth and gradual dependence on redshift. This form remains well-defined over a wide interval of redshift, extending even to the far-future regime $(z<0)$ and avoids divergences at high redshift. The logarithmic dependence ensures a continuous and scale-sensitive evolution of the deceleration parameter, naturally accommodating the transition from deceleration to acceleration without introducing unnecessary structural complexity. In addition, to explore the possible presence of non-trivial features in the expansion history, we also consider a log-periodic parametrization of the form $q(z)=q_0+q_1\sin[\log(1+z)]$~\cite{samaddar2025reconstructing}, which incorporates an oscillatory behavior on a logarithmic redshift scale. Such a formulation makes it possible to account for mild fluctuations or successive transitions between accelerating and decelerating phases during the evolution of the universe, which may arise due to effective gravitational corrections or underlying high-energy effects. Within the framework of $f(T)$ gravity, these parametrizations provide a convenient and model-independent approach to reconstruct the cosmic dynamics and examine possible deviations from the standard cosmological model. The logarithmic form offers a simple yet effective description of the expansion history, ensuring a smooth and continuous transition between different evolutionary phases of the universe. In contrast, the log-periodic parametrization, owing to its oscillatory nature, enables the exploration of potentially subtle features in the cosmic expansion dynamics. Such oscillatory behavior may be associated with variations in the effective equation of state or may indicate possible departures from standard gravitational theory during different stages of cosmic evolution. Therefore, the combined analysis of these parametrizations provides a flexible and physically motivated framework to probe the underlying nature of cosmic acceleration within modified teleparallel gravity.
\vspace{0.2cm}\\
This work is primarily devoted to examining an extended $f(T)$ gravity formulation developed in the framework of TEGR rather than the conventional curvature-based approach of GR. In this study, we employ parametrizations of the deceleration parameter and determine exact analytical solutions of the modified Friedmann equations for a homogeneous and isotropic FLRW universe. More precisely, we adopt two forms of $q(z)$, namely the logarithmic and log-periodic parametrizations, which allow for a flexible description of the cosmic expansion history. Furthermore, we utilize a power-law $f(T)$ gravity model characterized by $f(T)=\alpha(-T)^{n}$, with $\alpha$ and $n$ serving as free model parameters. The parameters of the adopted parametrizations are subsequently constrained using observational information from both the cosmic chronometer (CC) and combined (CC+Pantheon) datasets. A comprehensive investigation of the late-time cosmic evolution is carried out in the context of $f(T)$ gravity by exploring the evolution of different cosmological parameters, thereby providing a consistent and insightful assessment of the model's viability in explaining the observed late-time accelerated expansion.
\vspace{0.2cm}\\
The organization of the paper is as follows: Sections (\ref{sec:2}) and (\ref{sec:3}) are devoted to the mathematical formulation of $f(T)$ gravity and the background cosmological dynamics of a spatially flat FLRW universe, respectively, thereby establishing the theoretical foundation for the subsequent analysis. In Section (\ref{sec:4}), we introduce the parametric representations of the deceleration parameter and the corresponding Hubble parameter through two distinct cosmological models, enabling a detailed study of the cosmic expansion history. Section (\ref{sec:5}) presents the Bayesian MCMC constraints on the model parameters obtained from the cosmic chronometer (CC) and combined (CC+Pantheon) observational datasets. The physical and dynamical behavior of the models is discussed in Section (\ref{sec:6}), where we systematically study the behavior of fundamental cosmological quantities such as energy density, pressure and the DE equation of state (EoS). In addition, we examine the validity of energy conditions, perform statefinder and Om$(z)$ analyses, investigate thermodynamic consistency and the evolution of entropy, and estimate the age of the universe for both models. To conclude, Section (\ref{sec:7}) offers a summary of the main results of this study along with the final conclusions.
%%%%%%%%%%%%%%%%%%%%%%%%%%%%%%%%%%%%%%%%%%%%%%%%%%%%%%%%%%%%%%%%%%%%%%%%%%%%
\section{Mathematical framework of $f(T)$ gravity theory}\label{sec:2}
This section outlines the fundamental mathematical formalism of $f(T)$ gravity. Based on the torsion scalar, $f(T)$ gravity represents a modified theory whose geometric action is determined by a function of algebraic form involving torsion. The line element is given by
\begin{equation}{\label{1}}
ds^{2}= g_{\mu\nu}dx^{\mu}dx^{\nu}= \eta_{lm}\theta^{l}\theta^{m},
\end{equation}
with the components
\begin{equation}{\label{2}}
dx^{\mu} = e^{\mu}_{l}\theta^{l},~~~~~  \theta^{l}= e^{l}_{\mu} dx^{\mu},
\end{equation}
where $\eta_{lm}$ = $\mathrm{diag}(-1,1,1,1)$ represents the metric associated with Minkowskian spacetime, and $\left\{e^{l}_{\mu}\right\}$ are the tetrad components. These tetrads obey the following relations
\begin{equation}{\label{3}}
e^{~~\mu}_{l} e^{l}_{~~\nu}= \delta^{\mu}_{\nu},~~~~~  e^{~~l}_{\mu} e^{\mu}_{~~m}= \delta^{l}_{m}.
\end{equation}
Within $f(T)$ gravity, the geometric framework is described by the Weitzenböck connection~\cite{aldrovandi2012teleparallel}, defined as
\begin{equation}{\label{4}}
	\Gamma^{\alpha}_{\mu \nu}= e_{l}^{~\alpha} \partial_{\mu} e^{l}_{~\nu}= -e^{l}_{~\mu}\partial_{\nu} e^{~\alpha}_{l}.
\end{equation}
Using this connection, the torsion tensor components can be written as~\cite{linder2010einstein}
\begin{equation}{\label{5}}
T^{\alpha}_{~\mu \nu} = -\left(\Gamma^{\alpha}_{\nu \mu}-\Gamma^{\alpha}_{\mu \nu}\right)= - e^{~\alpha}_{l} \left(\partial_{\mu} e^{l}_{~\nu} - \partial_{\nu} e^{l}_{~\mu}\right).
\end{equation}
This tensor plays a central role in defining the contorsion tensor, which can be written as
\begin{equation}{\label{6}}
K^{\mu \nu}_{~\alpha} = -\frac{1}{2} \left(T^{\mu \nu}_{~\alpha} - T^{\nu \mu}_{\alpha} - T^{~\mu \nu}_{\alpha}\right),
\end{equation}
which, along with the torsion tensor, defines the superpotential tensor
\begin{equation}{\label{7}}
S^{~\mu \nu}_{\alpha} = \frac{1}{2} \left(K^{\mu \nu}_{\alpha} + \delta^{\mu}_{\alpha} T^{\lambda \nu}_{~\lambda} - \delta^{\nu}_{\alpha}T^{\lambda \mu}_{~\lambda}\right).
\end{equation}
The torsion scalar $T$, built from the torsion tensor and the superpotential $S^{~\mu \nu}_{\alpha}$, and is expressed as~\cite{cai2016f,maluf2013teleparallel}
\begin{equation}{\label{8}}
T= S^{~\mu \nu}_{\alpha} T^{\alpha}_{~\mu \nu} = \frac{1}{2}T^{\alpha \mu \nu } T_{\alpha \mu \nu} + \frac{1}{2}T^{\alpha \mu \nu } T_{\nu \mu \alpha} - T^{~\alpha}_{\alpha \mu } T^{\nu \mu}_{~\nu}.
\end{equation} 
Furthermore, the action for this gravitational theory can be written as~\cite{Bengochea,koussour2024exploring}
\begin{equation}{\label{9}}  
	S=  \frac{1}{2\kappa^{2}}\int d^{4}xe \left[T+f(T)\right] + \int d^{4}xe L_{m},
\end{equation}
where, $e$ refers to the determinant associated with the tetrad, given by $e= det (e^{l}_{~\mu}) = \sqrt{-g}$. The field equations of $f(T)$ gravity follow from the variation of the action~(\ref{9}) with respect to the tetrad fields and are given by:
\begin{equation}{\label{10}}
S^{~\nu \rho}_{\mu} \partial_{\rho} T f_{TT} + [e^{-1} e^{l}_{\mu} \partial_{\rho} (ee^{~\mu}_{l} S^{~\nu \lambda}_{\alpha}) + T^{\alpha}_{~\lambda \mu} S^{~\nu \lambda}_{\alpha}] f_{T} + \frac{1}{4}\delta^{\nu}_{\mu} f = \frac{\kappa^{2}}{2} \mathit{T}_{\mu}^{\nu},
\end{equation}
where $f_{T}= \frac{\partial f}{\partial T}$, $f_{TT}= \frac{\partial^{2} f}{\partial T^{2}}$ and $\mathit{T}_{\mu}^{\nu}$ corresponds to the energy-momentum tensor, expressed as
\begin{equation}{\label{11}}
	\mathit{T}_{\mu}^{\nu} = (\rho + \mathit{p}) u_{\mu} u^{\nu} + p \delta^{\nu}_{\mu},
\end{equation}
where $p$ and $\rho$ denote the pressure and energy density of the matter content of the universe, respectively. The four-velocity field $u^{\mu}$ obeys the normalization condition $u^{\mu}u_{\mu}=-1$.
\vspace{0.1cm}\\
The subsequent section derives the equations of motion for the FLRW cosmological background.
%%%%%%%%%%%%%%%%%%%%%%%%%%%%%%%%%%%%%%%%%%%%%%%%%%%%%%%%%%%%%%%%%%%%%%
\section{Equations governing $f(T)$ cosmology}\label{sec:3}
In this work, we consider a spatially flat FLRW metric, which is typically employed to implement the aforementioned theory within a cosmological context. This assumption facilitates the derivation of the modified Friedmann equations. The line element corresponding to a flat FLRW spacetime takes the form~\cite{linder2010einstein}
\begin{equation}{\label{12}}
	ds^{2}=-dt^{2}+a^{2}(t) \delta_{lm} dx^{l} dx^{m},
\end{equation}
where $a(t)$ denotes the scale factor. Consequently, the torsion scalar can be evaluated for the line element~(\ref{12}) as $T=-6H^{2}$.
\vspace{.1cm}\\
The Friedmann equations corresponding to the metric (\ref{12}) are given by~\cite{cai2016f}:
\begin{equation}{\label{13}}
6H^{2}+ 12H^{2}f_{T}+f = 2 \kappa^{2}\rho,
\end{equation}
\begin{equation}{\label{14}}
2\left(2\dot{H}+3H^{2}\right)+f+4 \left(\dot{H}+3H^{2}\right) f_{T}-48H^{2}\dot{H}f_{TT}=-2 \kappa^{2}p.
\end{equation}
Here, an overdot denotes differentiation with respect to cosmic time $t$, while $H$ denotes the Hubble parameter. The symbols $\rho$ and $p$ correspond to the energy density and pressure of the matter sector, respectively. Upon fixing $\kappa^{2}=1$, Eqs.~(\ref{13}) and (\ref{14}) can be recast as
\begin{equation}{\label{15}}
	3H^{2}= \rho + \rho_{DE},
\end{equation}
\begin{equation}{\label{16}}
-2 \dot{H}-3H^{2}= p+p_{DE}.
\end{equation}
Here, the energy density and pressure attributed to DE are given as follows
\begin{equation}{\label{17}}
	\rho_{DE}= -6H^{2}f_{T}-\frac{1}{2}f,
\end{equation}
\begin{equation}{\label{18}}
	p_{DE}= \frac{1}{2}f + 2\left(\dot{H}+3H^{2}\right) f_{T}-24\dot{H} H^{2} f_{TT}.
\end{equation}
From Eqs.~(\ref{17}) and (\ref{18}), we obtain the expression of the EoS parameter of DE as
\begin{equation}{\label{19}}
\omega_{DE}=\frac{p_{DE}}{\rho_{DE}} = -1-\frac{4\dot{H} \left(f_{T}-12H^{2}f_{TT}\right)}{f+ 12H^{2}f_{T}}.
\end{equation}
The above expression for the DE EoS parameter ($\omega_{DE}$), written in terms of the Hubble parameter $H(z)$ and the chosen $f(T)$ function, provides a convenient relation for investigating the expansion dynamics of the universe. This relation allows us to examine the evolution of the cosmic equation of state under different cosmological scenarios through appropriate choices of the deceleration parameter $q(z)$. In the following section, we introduce specific parametrized forms of $q(z)$ and explore their cosmological implications.
%%%%%%%%%%%%%%%%%%%%%%%%%%%%%%%%%%%%%%%%%%%%%%%%%%%%%%%%%%%%
\section{Parametric representations of the deceleration parameter}\label{sec:4}
\begin{figure}[ht]
	\centering
	\includegraphics[width=13cm, height=6cm]{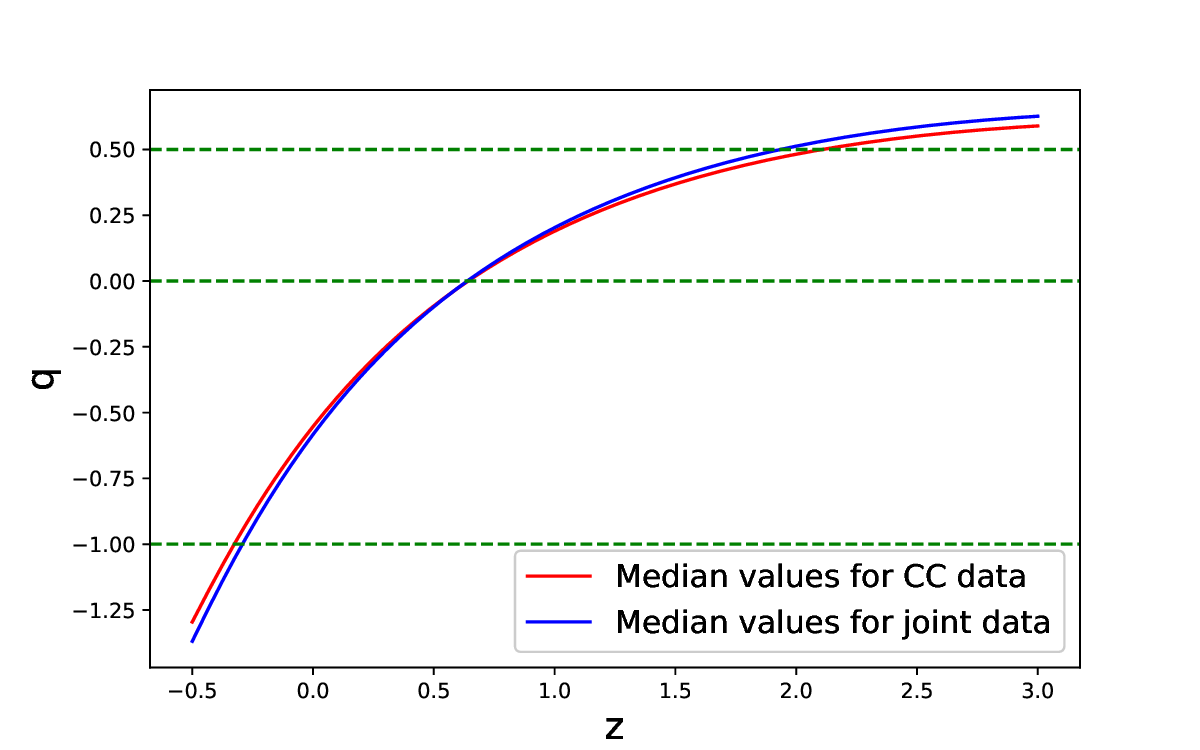}
	\caption{\textbf{For Model-2:} Plot of $q(z)$ with $z$.}
	\label{fig:2}
\end{figure}

The deceleration parameter $q$ is a fundamental quantity that governs the expansion dynamics of the universe. It determines the nature of cosmic evolution, where $q<0$ corresponds to an accelerating phase of expansion, while $q>0$ indicates decelerating behavior. For $q<-1$, the universe enters a super-accelerated (phantom) regime. Specific values of $q$ are associated with different cosmological epochs; for instance, $q=-1$ corresponds to the de Sitter phase, $q=\tfrac{1}{2}$ represents the matter-dominated era, and $q=1$ describes the radiation-dominated era. In this regard, a variety of studies have investigated both parametric and non-parametric descriptions of $q$. Such approaches have been extensively utilized to address several fundamental challenges in cosmology, including the initial singularity problem, persistent decelerating models, the horizon problem and the Hubble tension, among other key issues~\cite{banerjee2005acceleration,cunha2008transition,escamilla2022dynamical}. Consequently, various functional forms of $q(z)$ have been extensively investigated in the literature~\cite{chaudhary2025extracting,arora2024diagnostic,gadbail2022parametrization,koussour2023modeling,shekh2024late,al2016divergence}.
\vspace{0.2cm}\\
In light of these motivations, the present study adopts two distinct functional forms of the deceleration parameter in terms of the redshift $z$, namely the logarithmic and log-periodic models, to describe the dynamics of the universe within the framework of $f(T)$ gravity. The logarithmic parametrization is given by $q(z)=q_0 + q_1 \log(1+z)$~\cite{sofuouglu2026scalar}, while the log-periodic parametrization is expressed as $q(z)=q_0 + q_1 \sin[\log(1+z)]$~\cite{samaddar2025reconstructing}. Here, $q_0$ denotes the present-day value of the deceleration parameter, whereas $q_1$ governs its evolution. These functional forms provide a flexible and well-motivated description of the cosmic expansion history, enabling a detailed study of the evolution of the deceleration parameter. The time derivative of the Hubble parameter satisfies the relation $\dot{H}=-(1+q(z))H^2$. Consequently, the Hubble parameter can be expressed in integral form as
\begin{equation}{\label{20}}
H(z)=H_0 \exp\left[\int_{0}^{z} (1+q(z))\, d\ln(1+z)\right].
\end{equation}
\textbf{Model-1}: We adopt the logarithmic model of the deceleration parameter given as~\cite{sofuouglu2026scalar}
\begin{equation}{\label{21}}
q(z)=q_0 + q_1 \log(1+z).
\end{equation}
Using Eqs.~(\ref{20}) and (\ref{21}), the Hubble parameter can be written in terms of the redshift $z$ as:
\begin{equation}{\label{22}}
H(z)=H_{0} (1+z)^{1+q_{0}} \exp\left[\frac{q_{1}}{2} \left(\log[1+z]\right)^{2} \right].
\end{equation} 
\textbf{Model-2}: Another adopted form is the log-periodic model for the deceleration parameter given as~\cite{samaddar2025reconstructing}
\begin{equation}{\label{23}}
q(z)=q_0 + q_1 \sin[\log(1+z)].
\end{equation}
By employing Eqs.~(\ref{20}) and (\ref{23}), the Hubble parameter can be expressed in terms of the redshift $z$ as: 
\begin{equation}{\label{24}}
H(z)=H_{0} (1+z)^{1+q_{0}} \exp\left[q_{1} \left(1-\cos\left(\log[1+z]\right)\right) \right],
\end{equation}
where $H_{0}$ denotes the present-day value of the Hubble parameter at $z=0$. The model parameters $H_{0}$, $q_{0}$ and $q_{1}$ are constrained using observational data for both models.
%%%%%%%%%%%%%%%%%%%%%%%%%%%%%%%%%%%%%%%%%%%%%%%%%%%%%%%%%%%
\begin{figure}[ht]
	\centering
	\includegraphics[width=13cm, height=6cm]{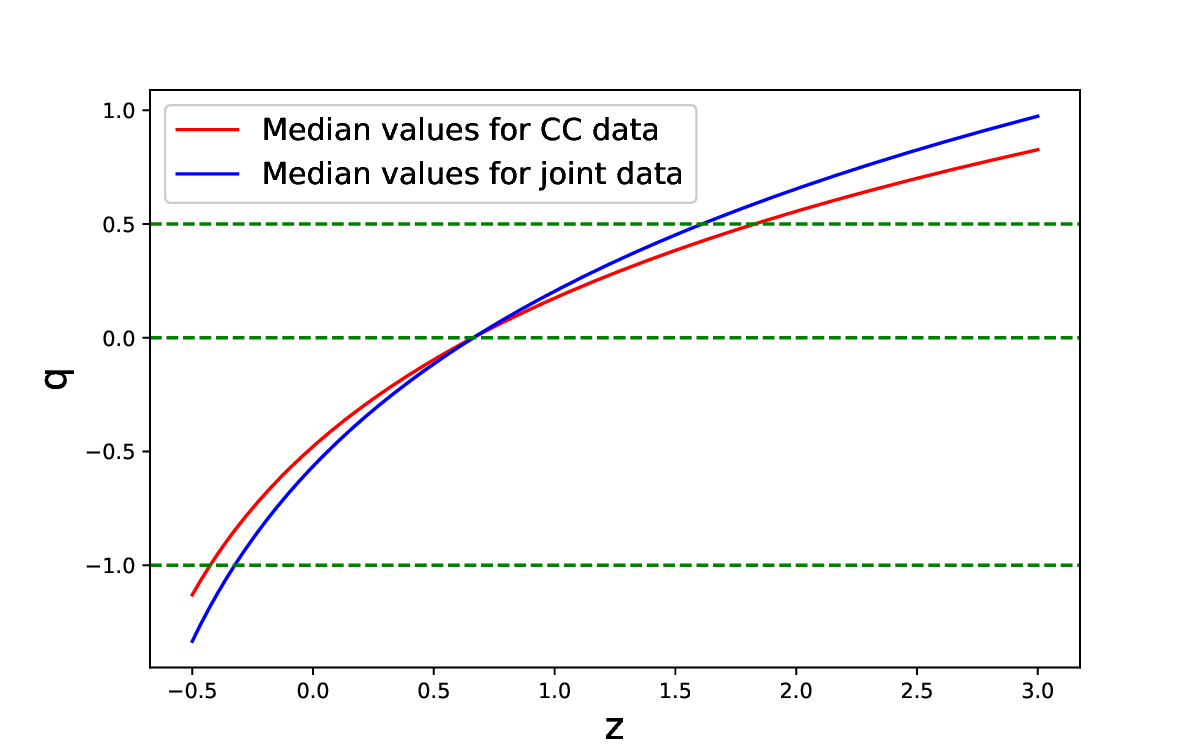}
	\caption{\textbf{For Model-1:} Plot of $q(z)$ with $z$.}
	\label{fig:1}
\end{figure}
%%%%%%%%%%%%%%%%%%%%%%%%%%%%%%%%%%%%%%%%%%%%%%%%%%%%%%%%%%%%%%%%%%%  
\vspace{0.2cm}\\
The analysis of the deceleration parameter $q(z)$ obtained from the CC dataset and the joint dataset is presented in Figures~(\ref{fig:1}) and (\ref{fig:2}), respectively. The results clearly indicate a transition of the universe from an earlier decelerating phase to the present accelerating epoch. Using the median values of the model parameters, the present-day value of the $q(z)$ for Model-1 is found to be $q_{0}=-0.478$ (CC dataset) and $q_{0}=-0.565$ (joint dataset). For Model-2, the corresponding values are $q_{0}=-0.553$ and $q_{0}=-0.583$ for the CC and joint datasets, respectively. The negative values of $q_{0}$ confirm that the universe is currently undergoing accelerated expansion at $z=0$. It is worth noting that the obtained values of $q_{0}$ are in close agreement with the observationally inferred $\Lambda$CDM prediction $q_{0} \approx -0.55$~\cite{2020A&A...641A...6P}. This consistency indicates that the present models successfully reproduce the observed late-time accelerated expansion behavior, in agreement with standard cosmological observations. At the same time, the models also allow for possible deviations from the standard $\Lambda$CDM scenario at future cosmic epochs, thereby accommodating more general dynamical evolutions such as phantom behavior. Moreover, the behavior of $q(z)$ consistently supports the existence of a transition from decelerated to accelerated expansion in the recent past. For Model-1, the transition redshift is obtained as $z=0.6692$ for both the CC and joint datasets, while for Model-2 it is $z=0.6563$ for both datasets.
%%%%%%%%%%%%%%%%%%%%%%%%%%%%%%%%%%%%%%%%%%%%%%%%%%%%%%%%%%%
\vspace{0.2cm}\\
The constraints obtained from both datasets indicate that the deceleration parameter approaches $q=\tfrac{1}{2}$ at intermediate redshifts, confirming that the universe undergoes a matter-dominated phase during its evolution for both models. As the universe evolves, a smooth transition to an accelerated expansion phase is observed, followed by a super-accelerated regime at late times driven by phantom-like DE. Notably, this late-time acceleration exceeds the de Sitter limit. Consequently, the effective DE component can be characterized as quintom in nature, consistent with the framework discussed in Ref.~\cite{Zhao2006}. This behavior demonstrates that the proposed models successfully capture the expected matter-dominated phase at earlier epochs while also accommodating the rich dynamics of late-time cosmic acceleration within a unified description.
%%%%%%%%%%%%%%%%%%%%%%%%%%%%%%%%%%%%%%%%%%%%%%%%%%%%%%%%%%%%%%%%%%%%%%%%%%%%%%% 
%%%%%%%%%%%%%%%%%%%%%%%%%%%%%%%%%%%%%%%%%%%%%%%%%%%%%%%%%%%
%%%%%%%%%%%%%%%%%%%%%%%%%%%%%%%%%%%%%%%%%%%%%%%%%%%%%%%%%%%       
%%%%%%%%%%%%%%%%%%%%%%%%%%%%%%%%%%%%%%%%%%%%%%%%%%%%%%%%%%%%%%%%%%%%%%%
\section{Observational constraints and results}\label{sec:5}
In this section, we employ a Bayesian inference approach to assess the consistency of the proposed cosmological models with observational datasets. The key model parameters, $H_{0}$, $q_{0}$ and $q_{1}$, appearing in the parametrized expressions of the Hubble parameter given in Eqs.~(\ref{22}) and (\ref{24}), are constrained using the cosmic chronometer (CC) dataset along with the joint (CC+Pantheon) dataset. The parameter estimation is carried out by minimizing the $\chi^{2}$ function within a Markov chain Monte Carlo (MCMC) framework, implemented using the emcee Python package~\cite{foreman2013emcee}. The resulting constraints are found to be statistically robust and consistent across both datasets, thereby reinforcing the reliability and predictive capability of the proposed models.
\subsection{The Cosmic chronometer dataset}\label{sec:5.1}
\begin{figure}[ht]
	\centering
	\includegraphics[width=13cm, height=6cm]{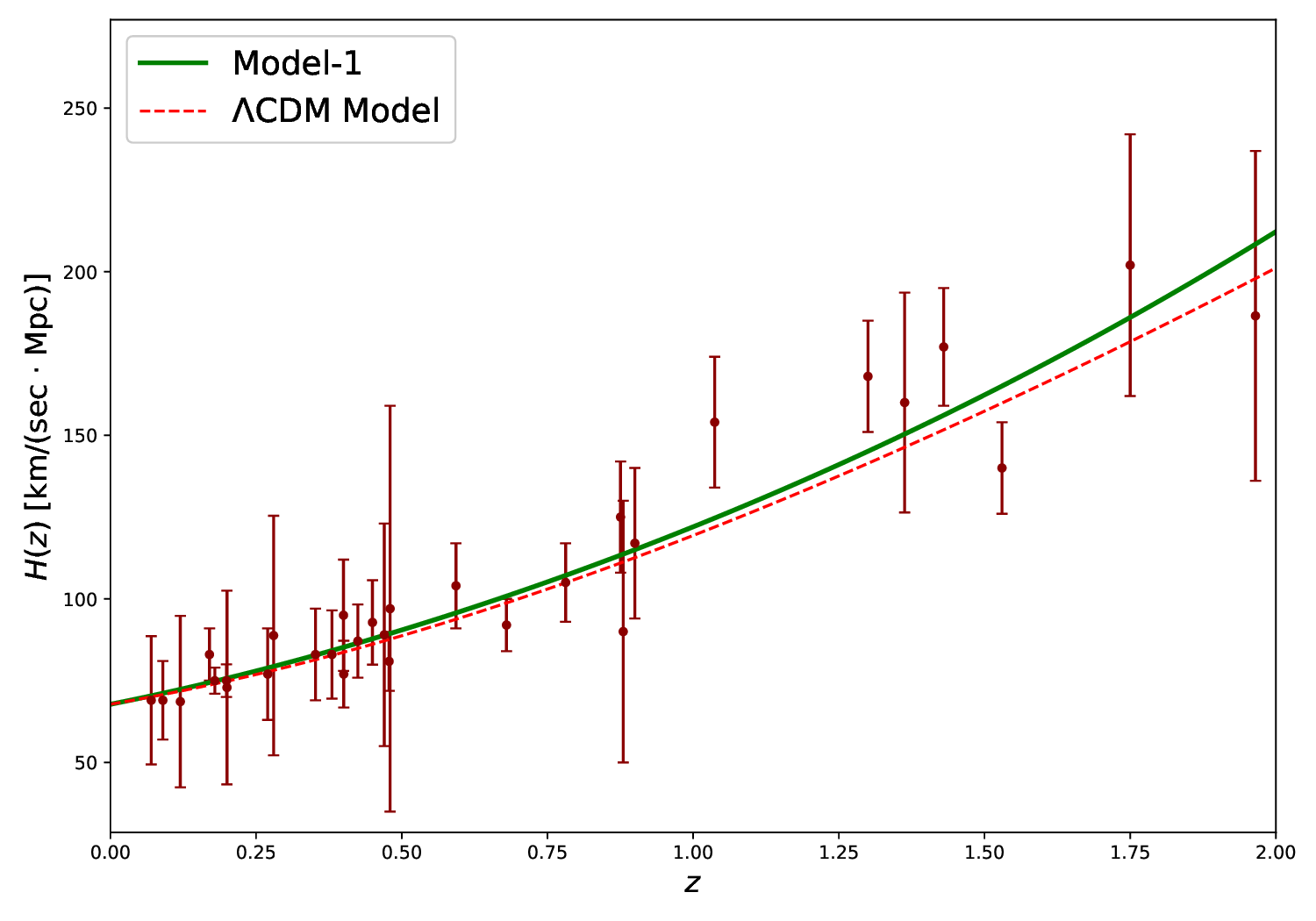}
	\caption{\textbf{For Model-1:} The best-fit $H(z)$ curve for the proposed model is compared with the $\Lambda CDM$ model.}
	\label{fig:3}
\end{figure}

\begin{figure}[ht]
	\centering
	\includegraphics[width=13cm, height=6cm]{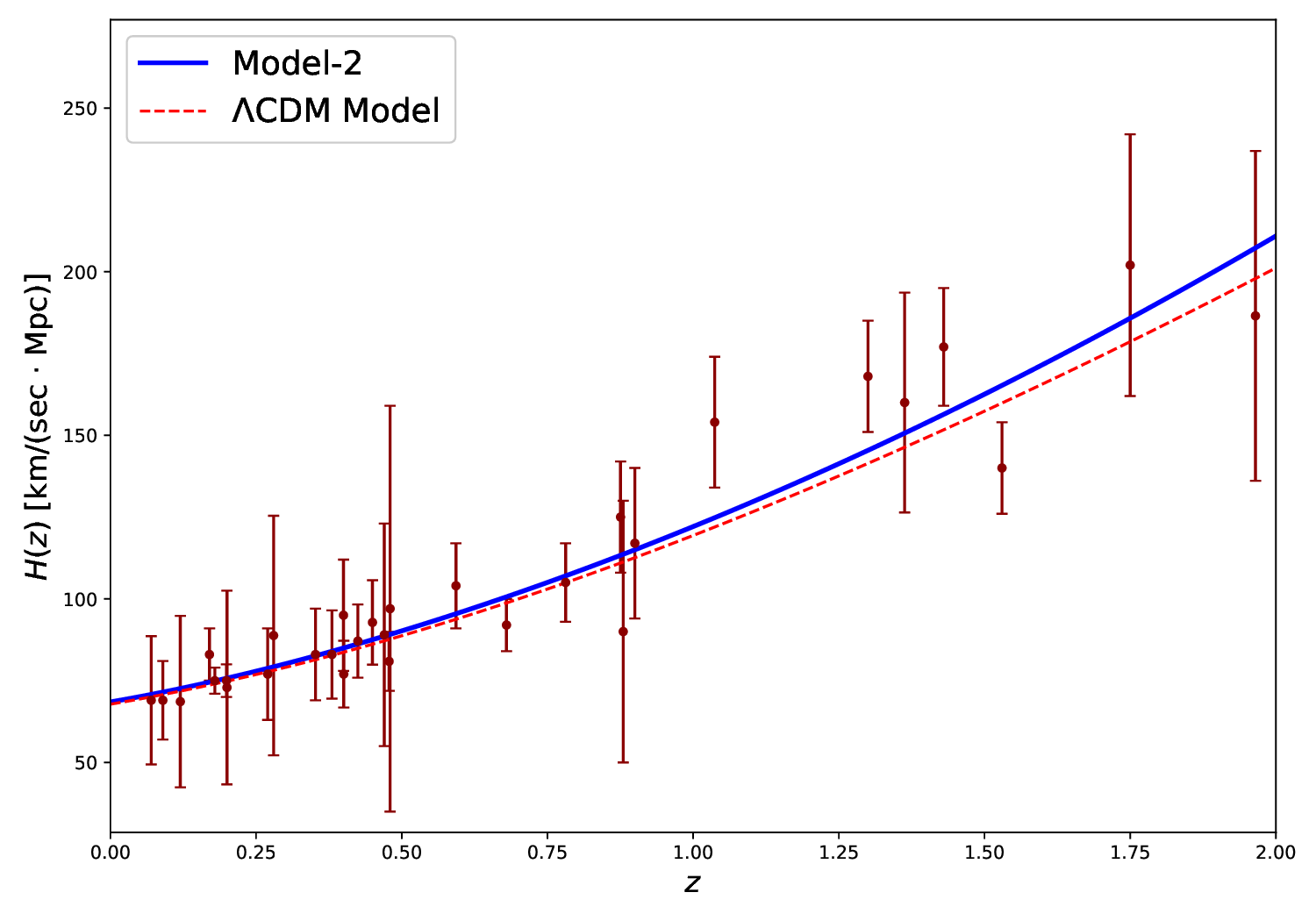}
	\caption{\textbf{For Model-2:} The best-fit $H(z)$ curve for the proposed model is compared with the $\Lambda CDM$ model.}
	\label{fig:4}
\end{figure}

To assess the compatibility of the proposed cosmological scenario with observations, we constrain the model parameters using available datasets. The analysis makes use of a compilation of $31$ cosmic chronometer (CC) measurements~\cite{simon2005constraints,sharov2018predictions}, obtained via the differential age (DA) technique applied to passively evolving galaxies across the redshift range $0.07 \leq z \leq 1.965$~\cite{stern2010cosmic,moresco2015raising}. The primary aim is to extract the best-fit values of the model parameters. Following the formalism introduced by Jimenez and Loeb~\cite{jimenez2002constraining}, the Hubble parameter is related to redshift $z$ and cosmic time $t$ through the relation $H(z) = -\frac{1}{(1+z)}\frac{dz}{dt}$. The free parameters $H_{0}$, $q_{0}$ and $q_{1}$ are then determined by minimizing the chi-squared $(\chi^{2})$ function (which is equivalent to maximizing the corresponding likelihood function) expressed as ~\cite{mandal2023cosmic,garg2025cosmological}:
\begin{equation}{\label{25}}
	\chi^{2}_{CC}(\theta)=\sum_{i=1}^{31} \frac{[H_{th}(\theta,z_{i})-H_{obs}(z_{i})]^{2} }{ \sigma^{2}_{H(z_{i})}},  
\end{equation} 
where $H_{\mathrm{th}}$ denotes the theoretical value of the Hubble parameter, $H_{\mathrm{obs}}$ represents the corresponding observed value and $\sigma_{H}$ indicates the associated uncertainty (standard deviation) of the observational data.
\vspace{0.1cm}\\
Figures~(\ref{fig:3}) and (\ref{fig:4}) illustrate the error bars associated with the CC data points, along with the corresponding best-fit Hubble parameter curves.
%%%%%%%%%%%%%%%%%%%%%%%%%%%%%%%%%%%%%%%%%%%%%%%%%%%%%%%%%%%
%%%%%%%%%%%%%%%%%%%%%%%%%%%%%%%%%%%%%%%%%%%%%%%%%%%%%%%%%%%
%%%%%%%%%%%%%%%%%%%%%%%%%%%%%%%%%%%%%%%%%%%%%%%%%%%%%%%%%%%
%%%%%%%%%%%%%%%%%%%%%%%%%%%%%%%%%%%%%%%%%%%%%%%%%%%%%%%%%%%              
%%%%%%%%%%%%%%%%%%%%%%%%%%%%%%%%%%%%%%%%%%%%%%%%%%%%%%%%%%%%%%%%%%%
\subsection{The Pantheon dataset}\label{sec:5.2}
\begin{center}
	\begin{figure}
		\includegraphics[width=19cm, height=19.5cm]{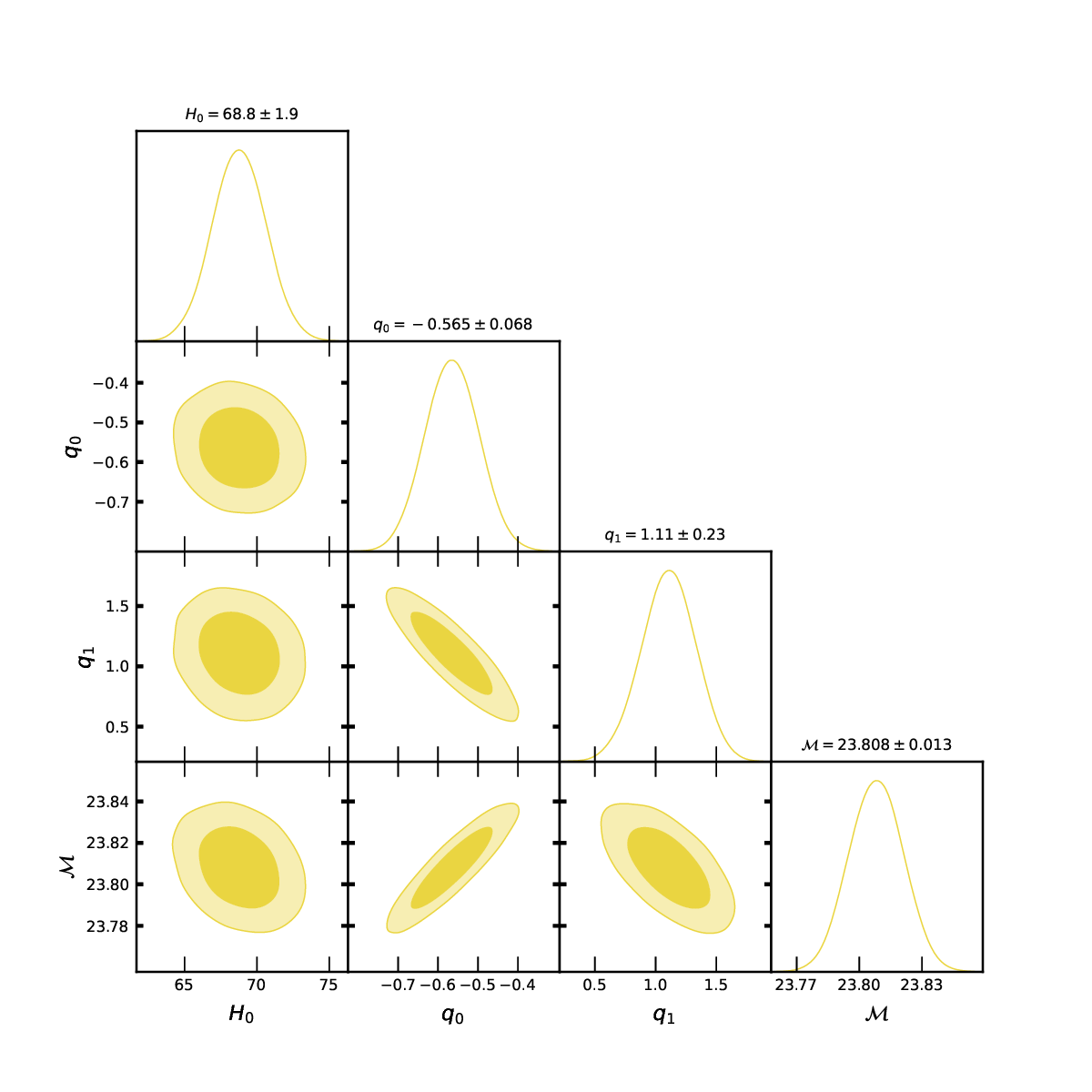}
		\caption{\textbf{For Model-1:} Marginalized $1D$ and $2D$ posterior contour map with median values of $H_{0}$, $q_{0}$ and $q_{1}$ using the Joint dataset.}
		\label{fig:5}
	\end{figure}
\end{center}

\begin{center}
	\begin{figure}
		\includegraphics[width=19cm, height=19.5cm]{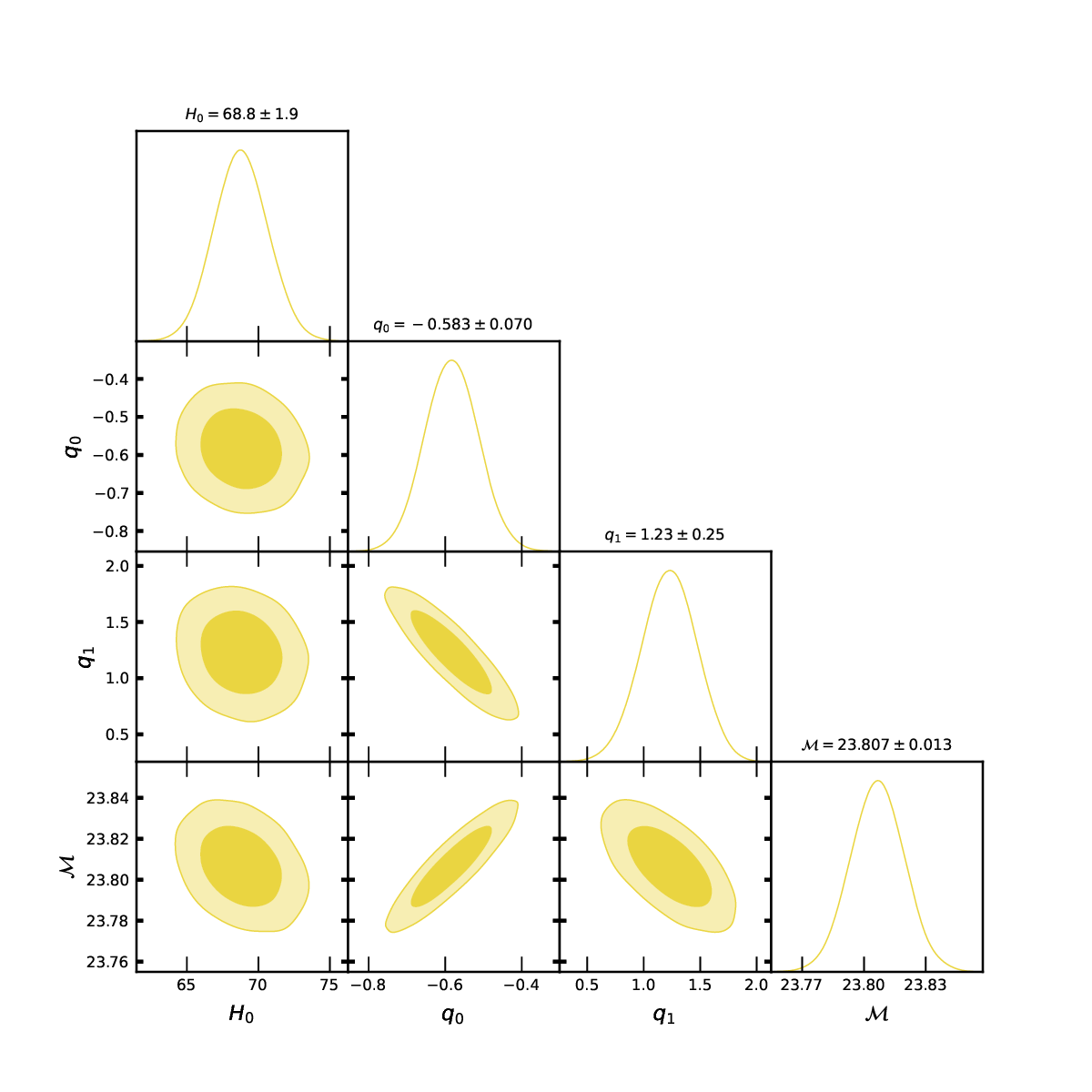}
		\caption{\textbf{For Model-2:} Marginalized $1D$ and $2D$ posterior contour map with median values of $H_{0}$, $q_{0}$ and $q_{1}$ using the Joint dataset.}
		\label{fig:6}
	\end{figure}
\end{center}

In the present analysis, we utilize the Pantheon compilation, which comprises 1048 Type Ia supernova (SNIa) data points spanning the redshift interval $0.01 < z < 2.26$~\cite{scolnic2018complete}. This extensive dataset integrates observations from several major surveys, including the CfA1–CfA4 series~\cite{riess1999bvri,hicken2009improved}, the Pan-STARRS1 Medium Deep Survey~\cite{scolnic2018complete}, SDSS~\cite{sako2018data}, SNLS~\cite{guy2010supernova} and the Carnegie Supernova Project (CSP)~\cite{contreras2010carnegie}. Within the MCMC framework applied to the Pantheon dataset, the theoretically predicted apparent magnitude $\mu_{\mathrm{th}}(z)$ is given by
\begin{equation}{\label{26}}
\mu_{th}(z)=25+5log_{10}\left[\frac{d_{L}(z)}{Mpc}\right]+M,
\end{equation}
here $M$ corresponds to the absolute magnitude, whereas the luminosity distance $d_{L}(z)$, possessing the dimension of length and is defined as~\cite{odintsov2018cosmological}
\begin{equation}{\label{27}}
	d_{L}(z)=c(1+z)\int_{0}^{z}\frac{dz'}{H(z')}.
\end{equation}
In this formulation, $z$ denotes the redshift of Type Ia supernovae (SNIa) measured in the cosmic microwave background (CMB) rest frame, while $c$ represents the speed of light. The luminosity distance $d_{L}(z)$ is often recast in terms of a dimensionless, Hubble-independent quantity defined as $D_{L}(z) \equiv H_{0}d_{L}(z)/c$. Consequently, equation~(\ref{26}) can be rewritten as follows
\begin{equation}{\label{28}}
	\mu_{th}(z)=25+5log_{10}\left[D_{L}(z)\right]+5log_{10}\left[\frac{c/H_{0}}{Mpc}\right]+M. 
\end{equation}
A known degeneracy between $M$ and $H_{0}$ is present in the $\Lambda$CDM framework~\cite{ellis2012relativistic,asvesta2022observational}. To mitigate this effect, we define a combined parameter $\mathcal{M}$ as follows
\begin{equation}{\label{29}}
	\mathcal{M}\equiv 25+5log_{10} \left[\frac{c/H_{0}}{Mpc}\right]+M=M+42.38-5log_{10}(h), 
\end{equation}
with $H_{0}=h \times 100$ $[\text{km}/(\text{sec}.\text{Mpc})]$, we carry out the MCMC analysis by incorporating these parameters along with the corresponding $\chi^{2}$ function for the Pantheon dataset, as defined by~\cite{asvesta2022observational,garg2025cosmological}:
\begin{equation}{\label{30}}
	\chi^{2}_{P}= \nabla \mu_{i}C^{-1}_{ij}\nabla \mu_{j}.
\end{equation}
Here $\nabla \mu_{i} = \mu_{obs}(z_{i}) - \mu_{th}(z_{i})$, where $C_{ij}^{-1}$ denotes the inverse covariance matrix and $\mu_{th}$ is given by equation (\ref{28}). The luminosity distance is inherently governed by the evolution of the Hubble parameter. In the present analysis, the emcee package~\cite{foreman2013emcee} is employed, in conjunction with the underlying theoretical framework, to perform maximum likelihood estimation (MLE) using the combined (CC+Pantheon) dataset. The joint $\chi^{2}$ function for the maximum likelihood estimate is constructed as the sum $\chi^{2}_{CC} + \chi^{2}_{P}$. Figures~(\ref{fig:5}) and (\ref{fig:6}) illustrate the $1\sigma$ and $2\sigma$ confidence regions in the form of contour plots, along with the corresponding 1D distributions obtained from the MCMC sampling of the joint CC+Pantheon datasets. The median values of the model parameters inferred from the MCMC constraints are summarized in Tables~(\ref{table:1}) and (\ref{table:2}).
%%%%%%%%%%%%%%%%%%%%%%%%%%%%%%%%%%%%%%%%%%%%%%%%%%%%%%%%%%%
%%%%%%%%%%%%%%%%%%%%%%%%%%%%%%%%%%%%%%%%%%%%%%%%%%%%%%%%%%%%%%%%%%%%%%
\begin{table}[htbp]
	\centering
	\renewcommand{\arraystretch}{2.5}  % Default 1.0 hai, isko badhayein
	\fontsize{10pt}{10pt}\selectfont   %{font size}{line spacing}
	\begin{tabular}{|c|c|c|c|c|c|c|c|c|c|}
			\hline
			Dataset & $H_{0}$[Km/(sec.Mpc)] & $q_{0}$ & $q_{1}$ & $\mathcal{M}$ & $z_{t}$ & $\omega_{0}$ & $t_{0}$(Gyr) \\
			\hline
			CC & $67.781^{+0.844}_{-0.846}$ & $-0.478^{+0.059}_{-0.060}$ & $0.941^{+0.112}_{-0.109}$ & - & $0.6692$ & $-0.6520$ & $12.73$ \\
			\hline
			CC+Pantheon  & $68.8^{+1.9}_{-1.9}$  & $-0.565^{+0.068}_{-0.068}$ &  $1.11^{+0.23}_{-0.23}$ &  $23.808^{+0.013}_{-0.013}$ & $0.6692$ & $-0.7100$ & $12.52$ \\
			\hline
	\end{tabular} 
     \caption{\textbf{For Model-1:} For both CC and joint datasets, the median values of the model parameters, along with the present values of $q_{0}$, $\omega_{0}$ and $t_{0}$.}
	\label{table:1}
\end{table}
%%%%%%%%%%%%%%%%%%%%%%%%%%%%%%%%%%%%%%%%%%%%%%%%%%%%%%%%%%%%%%%%%%%%%%
\begin{table}[htbp]
	\centering
	\renewcommand{\arraystretch}{2.5}  % Default 1.0 hai, isko badhayein
	\fontsize{10pt}{10pt}\selectfont   %{font size}{line spacing}
	\begin{tabular}{|c|c|c|c|c|c|c|c|c|c|}
		\hline
		Dataset & $H_{0}$[Km/(sec.Mpc)] & $q_{0}$ & $q_{1}$ & $\mathcal{M}$ & $z_{t}$ & $\omega_{0}$ & $t_{0}$(Gyr) \\
		\hline
		CC & $68.497^{+0.893}_{-0.888}$ & $-0.553^{+0.065}_{-0.065}$ & $1.162^{+0.129}_{-0.130}$ & - & $0.6563$ & $-0.7020$ & $14.61$ \\
		\hline
		CC+Pantheon  & $68.8^{+1.9}_{-1.9}$  & $-0.583^{+0.070}_{-0.070}$ &  $1.23^{+0.25}_{-0.25}$ &  $23.807^{+0.013}_{-0.013}$ & $0.6563$ & $-0.7220$ & $14.82$ \\
		\hline
	\end{tabular} 
	\caption{\textbf{For Model-2:} For both CC and joint datasets, the median values of the model parameters, along with the present values of $q_{0}$, $\omega_{0}$ and $t_{0}$.}
	\label{table:2}
\end{table}
%%%%%%%%%%%%%%%%%%%%%%%%%%%%%%%%%%%%%%%%%%%%%%%%%%%%%%%%%%%%
%%%%%%%%%%%%%%%%%%%%%%%%%%%%%%%%%%%%%%%%%%%%%%%%%%%%%%%%%%%%
%%%%%%%%%%%%%%%%%%%%%%%%%%%%%%%%%%%%%%%%%%%%%%%%%%%%%%%%%%%%%%%%%%%%
%%%%%%%%%%%%%%%%%%%%%%%%%%%%%%%%%%%%%%%%%%%%%%%%%%%%%%%%%%%%
%%%%%%%%%%%%%%%%%%%%%%%%%%%%%%%%%%%%%%%%%%%%%%%%%%%%%%%%%%%
%%%%%%%%%%%%%%%%%%%%%%%%%%%%%%%%%%%%%%%%%%%%%%%%%%%%%%%%%%%
\begin{figure}[!htb]
	\captionsetup{skip=0.4\baselineskip,size=footnotesize}
	\begin{minipage}{0.50\textwidth}
		\centering
		\includegraphics[width=8.6cm,height=7cm]{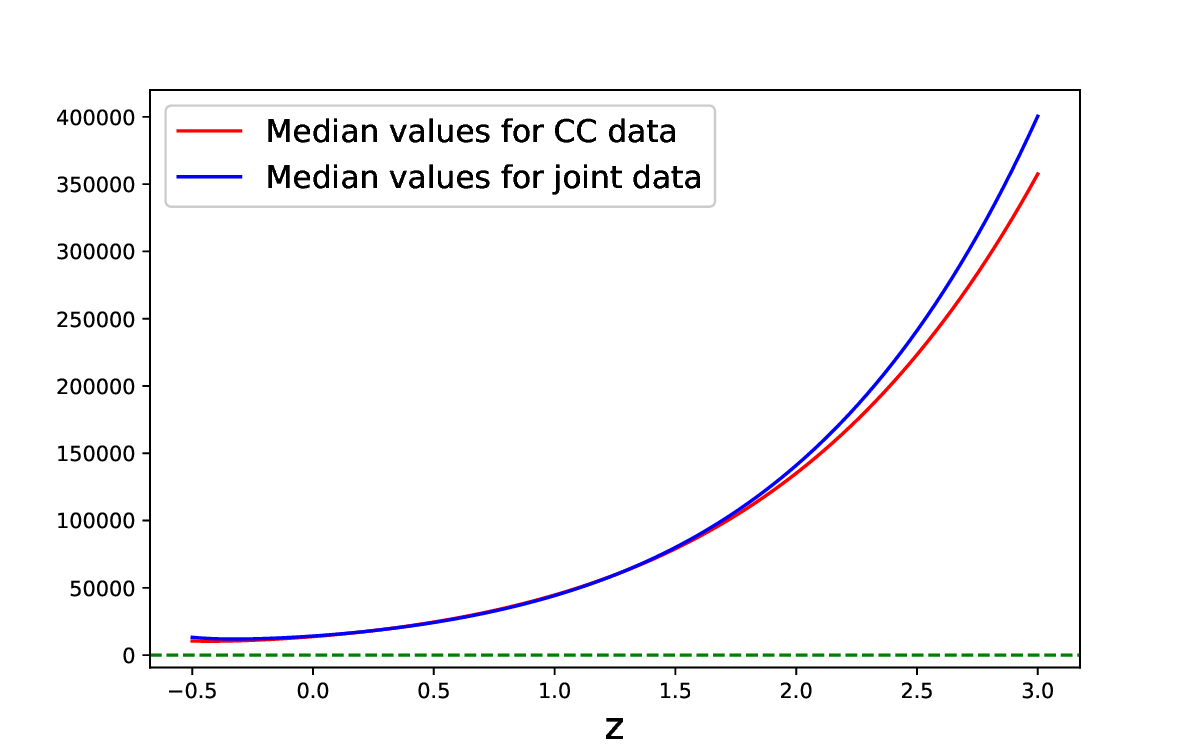}
		\caption{\textbf{For Model-1:} Plot of energy density ($\rho_{DE}$) with $\mathit{z}$.}
		\label{fig:7}
	\end{minipage}\hfill
	\begin{minipage}{0.50\textwidth}
		\centering
		\includegraphics[width=8.6cm,height=7cm]{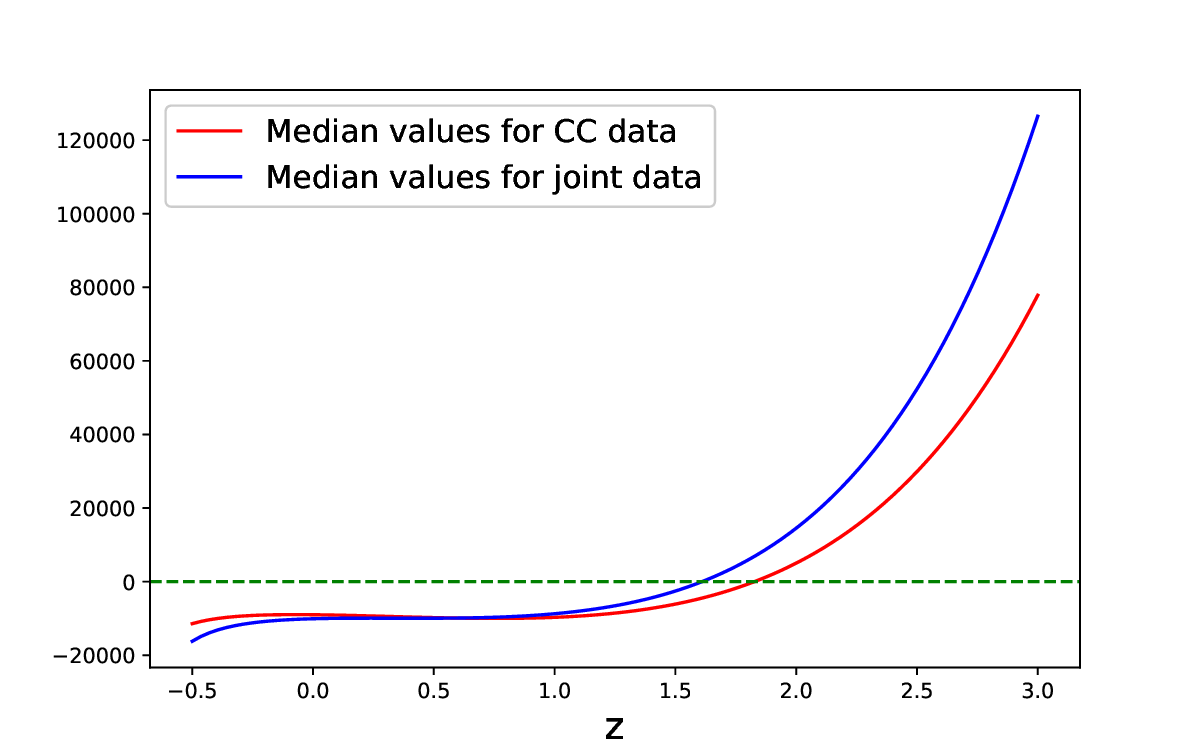}
		\caption{\textbf{For Model-1:} Plot of pressure ($p_{DE}$) with $\mathit{z}$.}
		\label{fig:8}
	\end{minipage}
\end{figure}
%%%%%%%%%%%%%%%%%%%%%%%%%%%%%%%%%%%%%%%%%%%%%%%%%%%%%%%%%%%%%%%
\section{The physical and dynamical characteristics}\label{sec:6}
%%%%%%%%%%%%%%%%%%%%%%%%%%%%%%%%%%%%%%%%%%%%%%%%%%%%%%%%%%%%%%%%%%%%%%%%%%
\subsection{Analyzing the physical behavior of the models}\label{sec:6.1}
\begin{figure}[!htb]
	\captionsetup{skip=0.4\baselineskip,size=footnotesize}
	\begin{minipage}{0.50\textwidth}
		\centering
		\includegraphics[width=8.6cm,height=7cm]{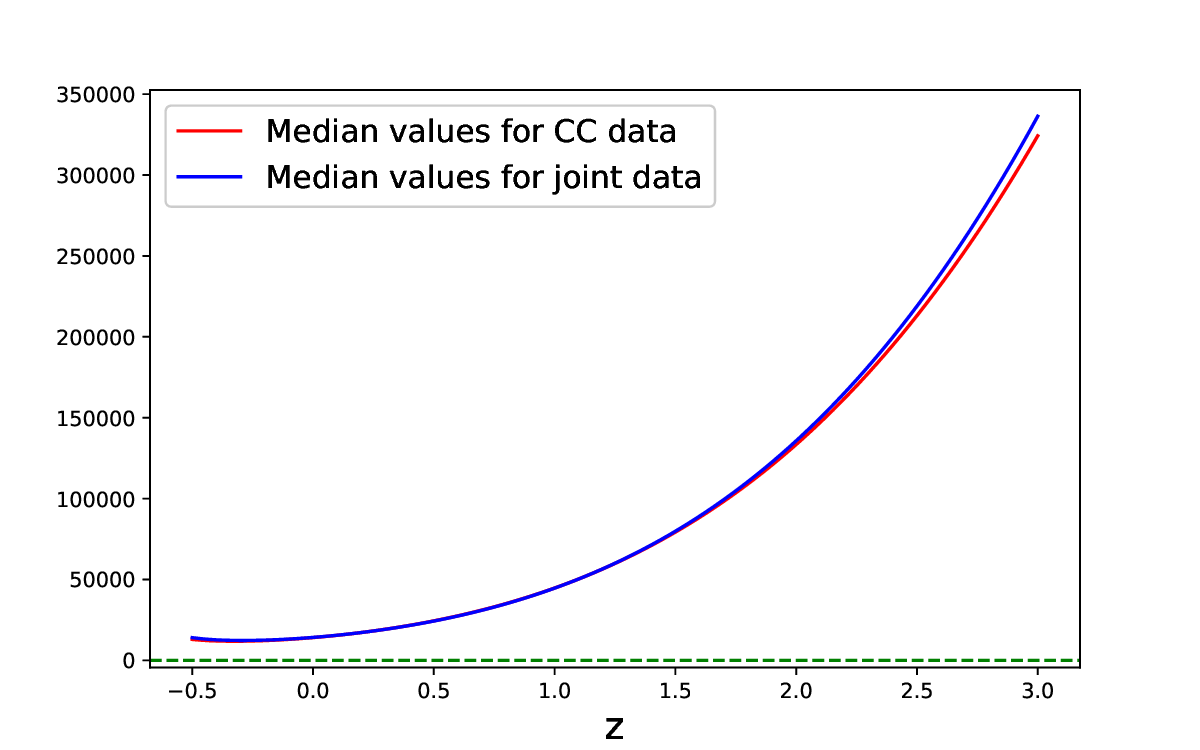}
		\caption{\textbf{For Model-2:} Plot of energy density ($\rho_{DE}$) with $\mathit{z}$.}
		\label{fig:9}
	\end{minipage}\hfill
	\begin{minipage}{0.50\textwidth}
		\centering
		\includegraphics[width=8.6cm,height=7cm]{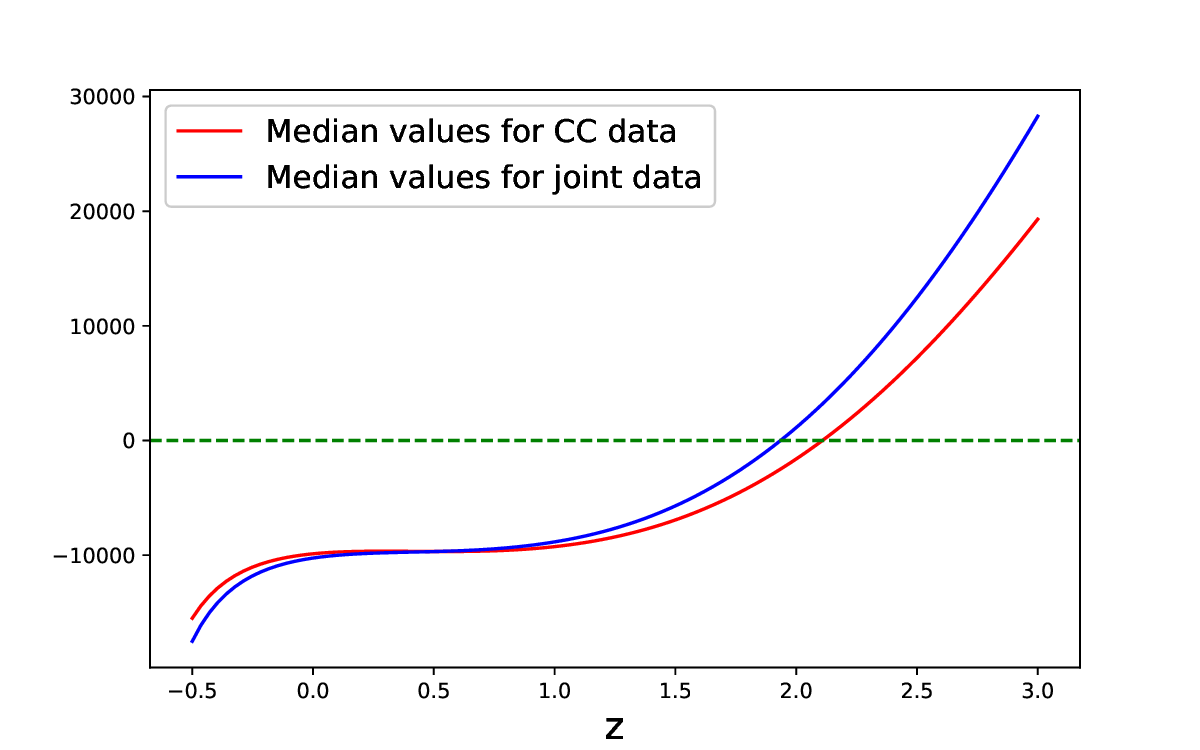}
		\caption{\textbf{For Model-2:} Plot of pressure ($p_{DE}$) with $\mathit{z}$.}
		\label{fig:10}
	\end{minipage}
\end{figure}

\begin{figure}[ht]
	\centering
	\includegraphics[width=13cm, height=6cm]{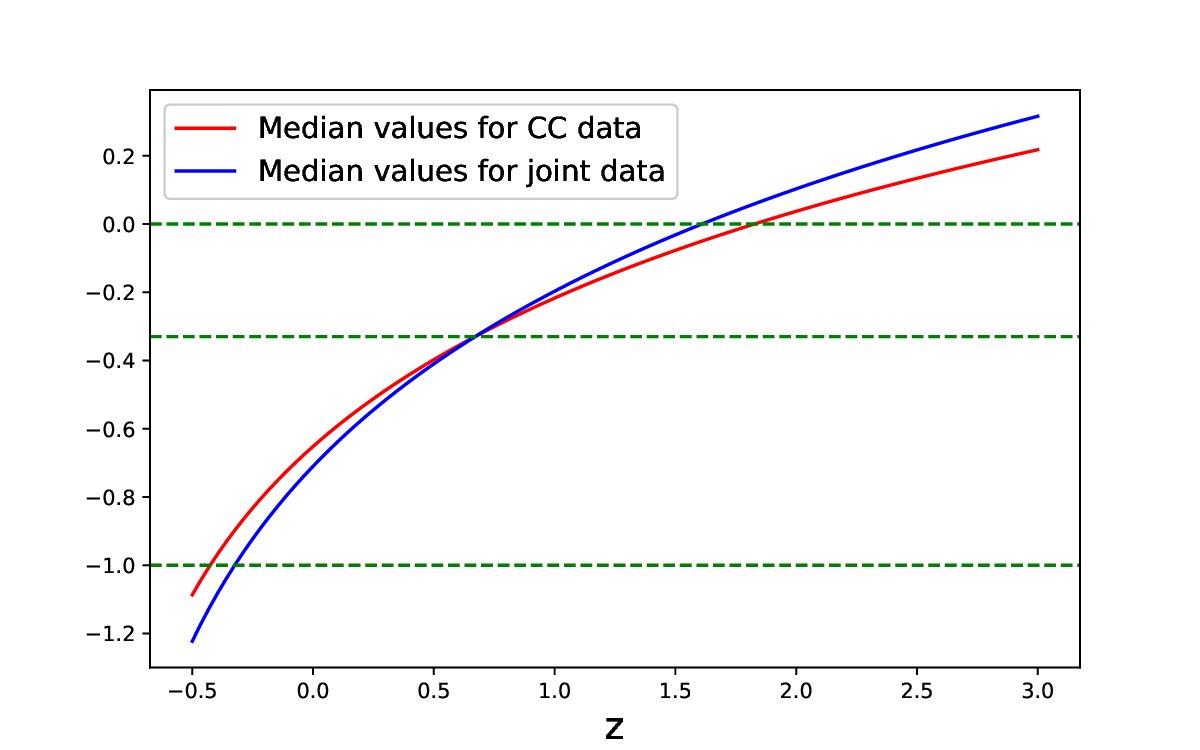}
	\caption{\textbf{For Model-1:} Plot of EoS parameter ($\omega_{DE}$) with $z$.}
	\label{fig:11}
\end{figure}

\begin{figure}[ht]
	\centering
	\includegraphics[width=13cm, height=6cm]{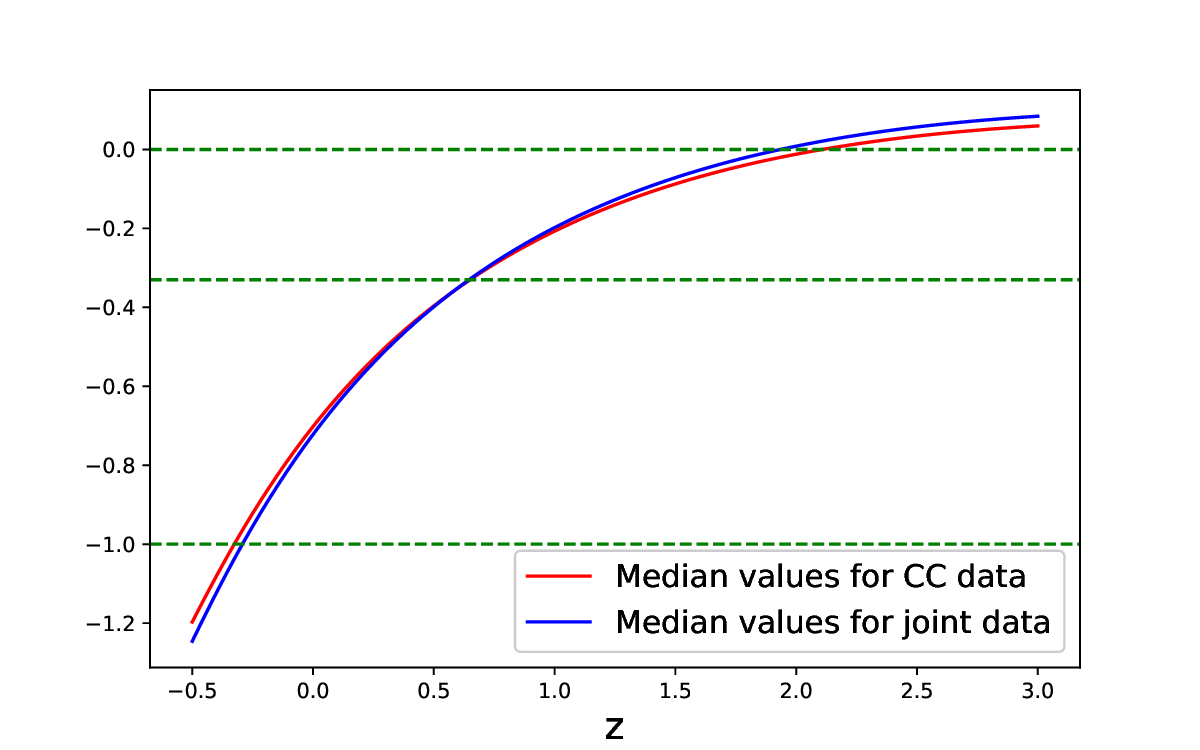}
	\caption{\textbf{For Model-2:} Plot of EoS parameter ($\omega_{DE}$) with $z$.}
	\label{fig:12}
\end{figure}

We analyze the physical characteristics of fundamental cosmological quantities, particularly the energy density and pressure. For the constrained set of model parameters, the energy density remains positive throughout the entire cosmic evolution, whereas the pressure exhibits a transition to negative values in the recent past. As the universe evolves from a decelerated phase to an accelerated expansion regime, the energy density preserves its positive nature, while the pressure attains negative values due to the increasing dominance of dark energy. This behavior is consistent with the expected features of a late-time accelerating universe. 
\vspace{0.2cm}\\
By using Eqs.~(\ref{17}), (\ref{18}) and (\ref{22}), the expressions for the DE density ($\rho_{\mathrm{DE}}$) and pressure ($p_{\mathrm{DE}}$) are derived as:
\begin{equation}{\label{31}}
\rho_{DE}(z) = \frac{(2n-1)}{2} \alpha 6^{n} H_{0}^{2n} (1+z)^{2n(1+q_{0})} \exp\left[n q_{1}\left(\log[1+z]\right)^{2}\right]            \qquad \qquad  \text{(for Model-1)},
\end{equation}
%%%%%%%%%%%%%%%%%%%%%%%%%%%%%%%%%%%%%%%%%%%%%%%%%%%%%%%%
\begin{equation} {\label{32}}
	\begin{split}
	p_{DE}(z)= & 2\alpha n 6^{n-1} H_{0}^{2n} \left(1+2(n-1)\right) (1+z)^{2n(1+q_{0})} \left[1+q_{0}+q_{1}\log(1+z)\right] \exp\left[n q_{1}\left(\log[1+z]\right)^{2}\right] \\& +  \alpha 6^{n} H_{0}^{2n} \left(\frac{1}{2}-n\right)  (1+z)^{2n(1+q_{0})} \exp\left[n q_{1}\left(\log[1+z]\right)^{2}\right]      \qquad \qquad     \text{(for Model-1)}.
	\end{split}
\end{equation} 
\vspace{0.2cm}\\
By employing Eqs.~(\ref{17}), (\ref{18}) and (\ref{24}), the expressions for the DE density ($\rho_{\mathrm{DE}}$) and pressure ($p_{\mathrm{DE}}$) are obtained as:
\begin{equation}{\label{33}}
	\rho_{DE}(z) = \frac{(2n-1)}{2} \alpha 6^{n} H_{0}^{2n} (1+z)^{2n(1+q_{0})} \exp\left[2n q_{1}\left(1-\cos\left(\log[1+z]\right)\right)\right]   \qquad \qquad  \text{(for Model-2)},
\end{equation}
%%%%%%%%%%%%%%%%%%%%%%%%%%%%%%%%%%%%%%%%%%%%%%%%%%%%%%%%
\begin{equation} {\label{34}}
	\begin{split}
		p_{DE}(z)= & 2\alpha n 6^{n-1} H_{0}^{2n} \left(1+2(n-1)\right) (1+z)^{2n(1+q_{0})} \left[1+q_{0}+q_{1}\sin\left(\log[1+z]\right)\right] \exp\left[2n q_{1}\left(1-\cos\left(\log[1+z]\right)\right)\right] \\& +  \alpha 6^{n} H_{0}^{2n} \left(\frac{1}{2}-n\right)  (1+z)^{2n(1+q_{0})} \exp\left[2n q_{1}\left(1-\cos\left(\log[1+z]\right)\right)\right]    \qquad\qquad \text{(for Model-2)}.
	\end{split}
\end{equation} 
\vspace{0.2cm}\\
The evolutionary behavior of the dark energy density ($\rho_{\mathrm{DE}}$) and pressure ($p_{\mathrm{DE}}$) is illustrated in Figures~(\ref{fig:7}) and (\ref{fig:8}) for Model-1 and in Figures~(\ref{fig:9}) and (\ref{fig:10}) for Model-2. For the observationally constrained values of the model parameters, the dark energy density ($\rho_{\mathrm{DE}}$) in both models exhibits a monotonic increase with redshift $z$ (corresponding to a decreasing behavior with cosmic time $t$), while remaining positive over the entire cosmic evolution. This positivity of $\rho_{\mathrm{DE}}$ ensures the physical viability of the model and confirms its role in driving the expansion of the universe. In contrast, the DE pressure ($p_{\mathrm{DE}}$) remains negative in both the present and future epochs. Such negative pressure's may lead to a more rapid expansion of the universe in the late-time regime. These features collectively indicate a dark energy-dominated phase of cosmic evolution and are consistent with the observed accelerating expansion of the universe. We take $\alpha=1$ and $n=1$ for these graphical illustrations. In cosmological studies, the Equation of State (EoS) parameter plays a central role in characterizing the physical properties of DE. It is defined as the ratio of pressure to energy density of the cosmic fluid ($\omega = \frac{p}{\rho}$) and serves as a key diagnostic for understanding the dynamical evolution of the universe. Distinct values of $\omega$ correspond to different cosmological epochs; specifically, $\omega = 0$ describes pressureless matter (dust), $\omega = \frac{1}{3}$ characterizes the radiation-dominated era and $\omega = -1$ represents vacuum energy associated with a de Sitter phase. A necessary condition for the onset of accelerated expansion is $\omega < -\frac{1}{3}$. This regime includes both the quintessence domain ($-1 < \omega < -\frac{1}{3}$) and the phantom domain ($\omega < -1$), which correspond to different dynamical realizations of DE and have distinct implications for the future evolution of the universe.
\vspace{0.2cm}\\
The dark energy EoS parameter ($\omega_{\mathrm{DE}}$), is derived from Eqs.~(\ref{31}) and (\ref{32}):
%%%%%%%%%%%%%%%%%%%%%%%%%%%%%%%%%%%%%%%%%%%%%%%%%%%%%%%%
\begin{equation} {\label{35}}
\omega_{DE}= -1+\frac{2}{3}n\left(1+q_{0}+q_{1} \log(1+z)\right) \qquad\qquad \qquad \text{(for Model-1)}
\end{equation} 
Further, the dark energy EoS parameter ($\omega_{\mathrm{DE}}$), is subsequently obtained from Eqs.~(\ref{33}) and (\ref{34}):
\begin{equation} {\label{36}}
	\omega_{DE}= -1+\frac{2}{3}n\left[1+q_{0}+q_{1} \sin\left(\log(1+z)\right)\right] \qquad\qquad \qquad \text{(for Model-2)}
\end{equation} 
%%%%%%%%%%%%%%%%%%%%%%%%%%%%%%%%%%%%%%%%%%%%%%%%%%%%%%%%%%%
%%%%%%%%%%%%%%%%%%%%%%%%%%%%%%%%%%%%%%%%%%%%%%%%%%%%%%%%%%%%%%%
The redshift evolution of the DE EoS parameter is depicted in Figures~(\ref{fig:11}) and (\ref{fig:12}) for the proposed models. At the present epoch ($z=0$), Model-1 predicts $\omega_{\mathrm{DE}}=-0.6520$ for the CC dataset and $\omega_{\mathrm{DE}}=-0.7100$ for the joint dataset. Correspondingly, Model-2 yields $\omega_{\mathrm{DE}}=-0.7020$ and $\omega_{\mathrm{DE}}=-0.7220$ for the CC and joint datasets, respectively. The evolution of $\omega_{\mathrm{DE}}(z)$ clearly indicates that, at the current epoch both models lie within the quintessence regime ($-1 < \omega_{\mathrm{DE}} < -\frac{1}{3}$) across the datasets considered. This result points toward a dynamically evolving DE component that effectively drives the late-time acceleration of the universe. Furthermore, the evolutionary trajectories suggest that, for the median parameter values, the EoS parameter undergoes a smooth transition across the cosmological constant boundary ($\omega_{DE} = -1$), entering the phantom regime at late times. Such a transition implies a progressively dominant DE component, leading to an enhanced rate of cosmic expansion and significantly influencing the future dynamics of the universe.
%%%%%%%%%%%%%%%%%%%%%%%%%%%%%%%%%%%%%%%%%%%%%%%%%%%%%%%
%%%%%%%%%%%%%%%%%%%%%%%%%%%%%%%%%%%%%%%%%%%%%%%%%%%%%%%%%%%
%%%%%%%%%%%%%%%%%%%%%%%%%%%%%%%%%%%%%%%%%%%%%%%%%%%%%%%%%%%
%%%%%%%%%%%%%%%%%%%%%%%%%%%%%%%%%%%%%%%%%%%%%%%%%%%%%%%%%%%
%%%%%%%%%%%%%%%%%%%%%%%%%%%%%%%%%%%%%%%%%%%%%%%%%%%%%%%%%%%              
%%%%%%%%%%%%%%%%%%%%%%%%%%%%%%%%%%%%%%%%%%%%%%%%%%%%%%%%%%%
\subsection{Analysis of energy conditions}\label{sec:6.2}
\begin{figure}[ht]
	\centering
	\includegraphics[width=13cm, height=6cm]{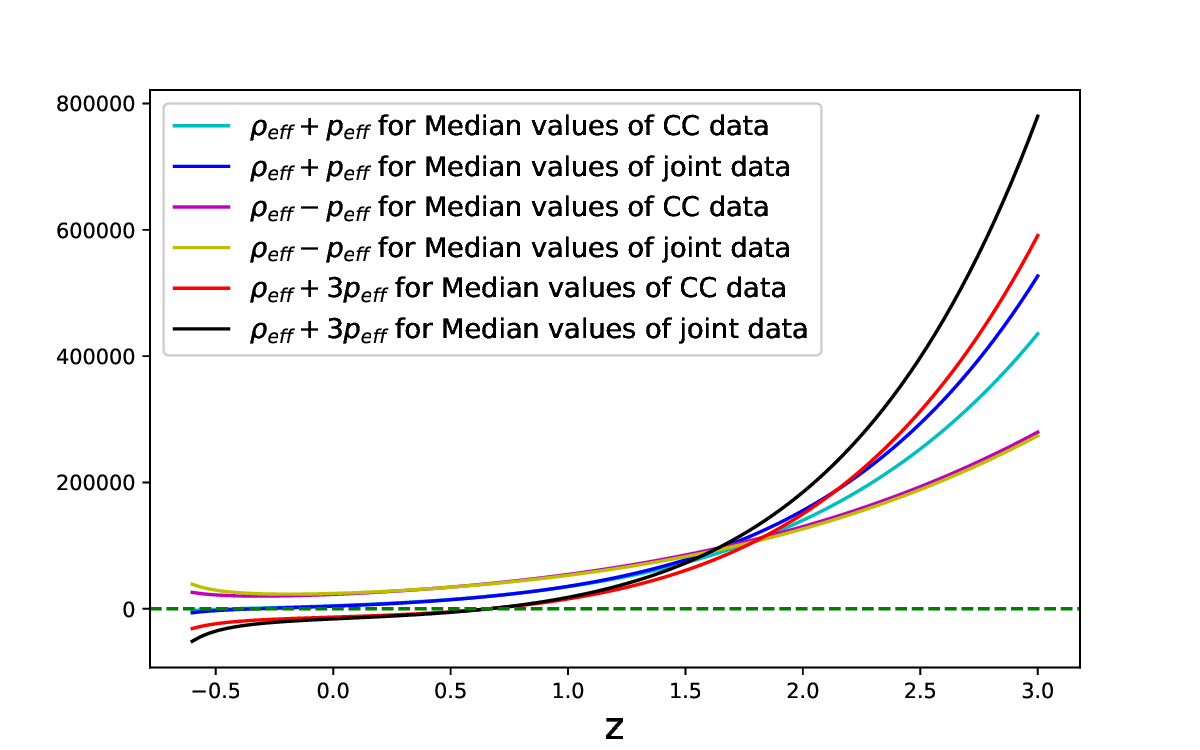}
	\caption{\textbf{For Model-1:} Plot of the components of energy conditions with $z$.}
	\label{fig:13}
\end{figure}

\begin{figure}[ht]
	\centering
	\includegraphics[width=13cm, height=6cm]{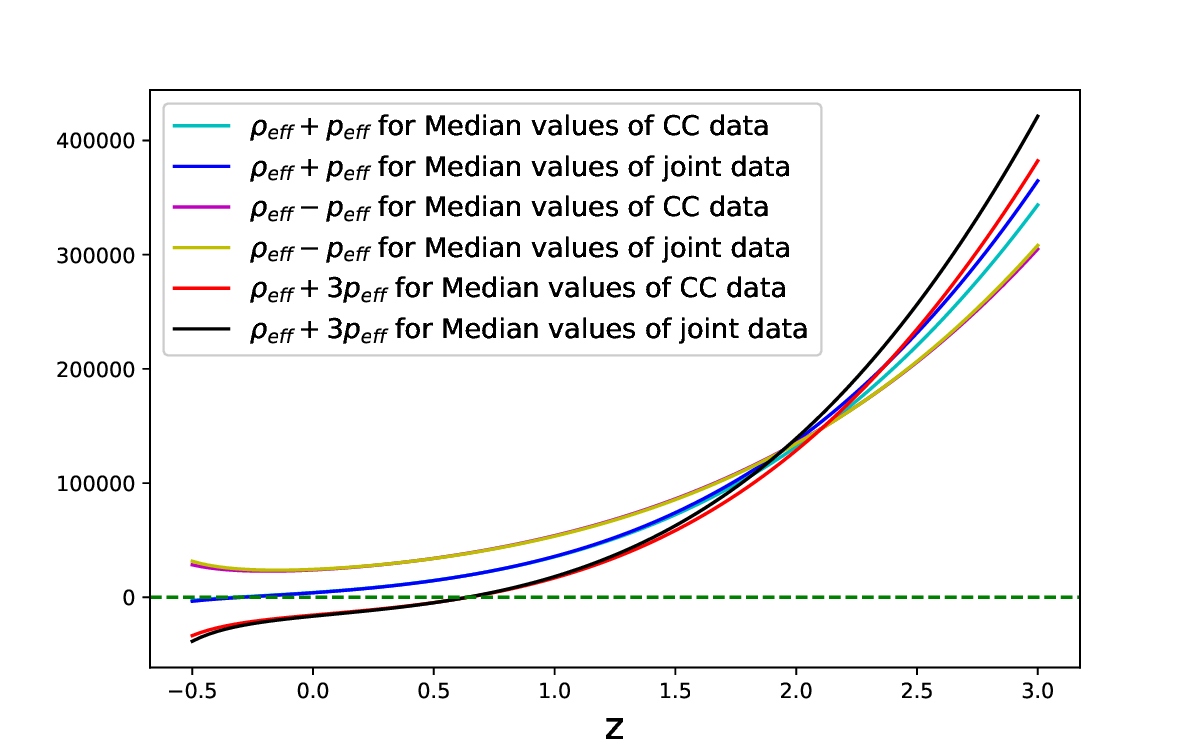}
	\caption{\textbf{For Model-2:} Plot of the components of energy conditions with $z$.}
	\label{fig:14}
\end{figure}

The point-wise energy conditions at a spacetime point, governed exclusively by the stress energy tensor, are expressed as follows~\cite{visser1997energy,lalke2024cosmic,singh2022lagrangian}:
%%%%%%%%%%%%%%%%%%%%%%%%%%%%%%%%%%%%%%%%%%%%%%%%%%%%%%%%%%%% 
\begin{itemize}
\item \textbf{NEC:-} The null energy condition requires that $\rho_{\mathrm{eff}} + p_{\mathrm{eff}} \geq 0$, ensuring that the combined contribution of the effective energy density and pressure remains non-negative.
	
\item \textbf{WEC:-} The weak energy condition is satisfied when $\rho_{\mathrm{eff}} \geq 0$ and $\rho_{\mathrm{eff}} + p_{\mathrm{eff}} \geq 0$, indicating that the effective energy density and its sum with pressure remain non-negative.
	
\item \textbf{DEC:-} The dominant energy condition requires that $\rho_{\mathrm{eff}} \geq |p_{\mathrm{eff}}|$, implying that the effective energy density is non-negative and greater than or equal to the magnitude of the pressure.
	
\item \textbf{SEC:-} The strong energy condition holds when $\rho_{\mathrm{eff}} + p_{\mathrm{eff}} \geq 0$ and $\rho_{\mathrm{eff}} + 3p_{\mathrm{eff}} \geq 0$, such that the effective energy density and pressure satisfy the required positivity conditions.
	
\end{itemize}
%%%%%%%%%%%%%%%%%%%%%%%%%%%%%%%%%%%%%%%%%%%%%%%%%%%%%%%%%%%%
The Strong Energy Condition (SEC) incorporates the inequality $\rho_{\mathrm{eff}} + 3p_{\mathrm{eff}} \geq 0$, which is fundamentally linked to the Raychaudhuri equation and governs the dynamical behavior of cosmic expansion~\cite{mishra2025cosmological}. In standard cosmological evolution, a positive value of the active gravitational mass term $(\rho_{\mathrm{eff}} + 3p_{\mathrm{eff}})$ corresponds to a decelerating universe. However, a growing body of observational evidence indicates that this condition is violated during the transition from the epoch of structure formation to the present accelerated phase. This violation characterized by $\rho_{\mathrm{eff}} + 3p_{\mathrm{eff}} < 0$, leads to a repulsive gravitational effect and thereby drives the late-time acceleration of the universe. Such behavior implies the existence of components with sufficiently negative pressure, consistent with the phenomenology of DE. It is important to note that the SEC consists of two independent inequalities and the violation of either condition is sufficient to signal a breakdown of the SEC~\cite{mishra2025cosmological}.
\vspace{0.2cm}\\
Figures~(\ref{fig:13}) and (\ref{fig:14}) present the evolution of the energy conditions for Model-1 and Model-2, respectively. The graphical analysis evident that both models satisfy the Null Energy Condition (NEC), Weak Energy Condition (WEC) and Dominant Energy Condition (DEC) up to the present epoch. However, the transition to an accelerated expansion phase necessitates the violation of the SEC (specifically the constraint $\rho_{eff} + 3p_{eff} \geq 0$). As the cosmic evolution progresses into the phantom regime, the NEC is found to be violated, which in turn can lead to the breakdown of the WEC and DEC. The violation ($\rho_{\mathrm{eff}} + p_{\mathrm{eff}} \geq 0$) serves as a clear signature of phantom-type dark energy behavior in both models. Consequently, these findings suggest that a phantom DE component cannot be ruled out in the present framework. Furthermore, the continued violation of $\rho_{\mathrm{eff}} + 3p_{\mathrm{eff}} \geq 0$ during the phantom-dominated era consistently supports a sustained phase of accelerated expansion driven by negative pressure. These results reinforce the consistency of the models with the EoS behavior and the late-time accelerated expansion.
%%%%%%%%%%%%%%%%%%%%%%%%%%%%%%%%%%%%%%%%%%%%%%%%%%%%%%%
%%%%%%%%%%%%%%%%%%%%%%%%%%%%%%%%%%%%%%%%%%%%%%%%%%%%%%%%%%%
%%%%%%%%%%%%%%%%%%%%%%%%%%%%%%%%%%%%%%%%%%%%%%%%%%%%%%%%%%%
%%%%%%%%%%%%%%%%%%%%%%%%%%%%%%%%%%%%%%%%%%%%%%%%%%%%%%%%%%%
%%%%%%%%%%%%%%%%%%%%%%%%%%%%%%%%%%%%%%%%%%%%%%%%%%%%%%%%%%%
%%%%%%%%%%%%%%%%%%%%%%%%%%%%%%%%%%%%%%%%%%%%%%%%%%%%%%%%%%%
\begin{figure}[ht]
	\centering
	\includegraphics[width=11cm, height=6cm]{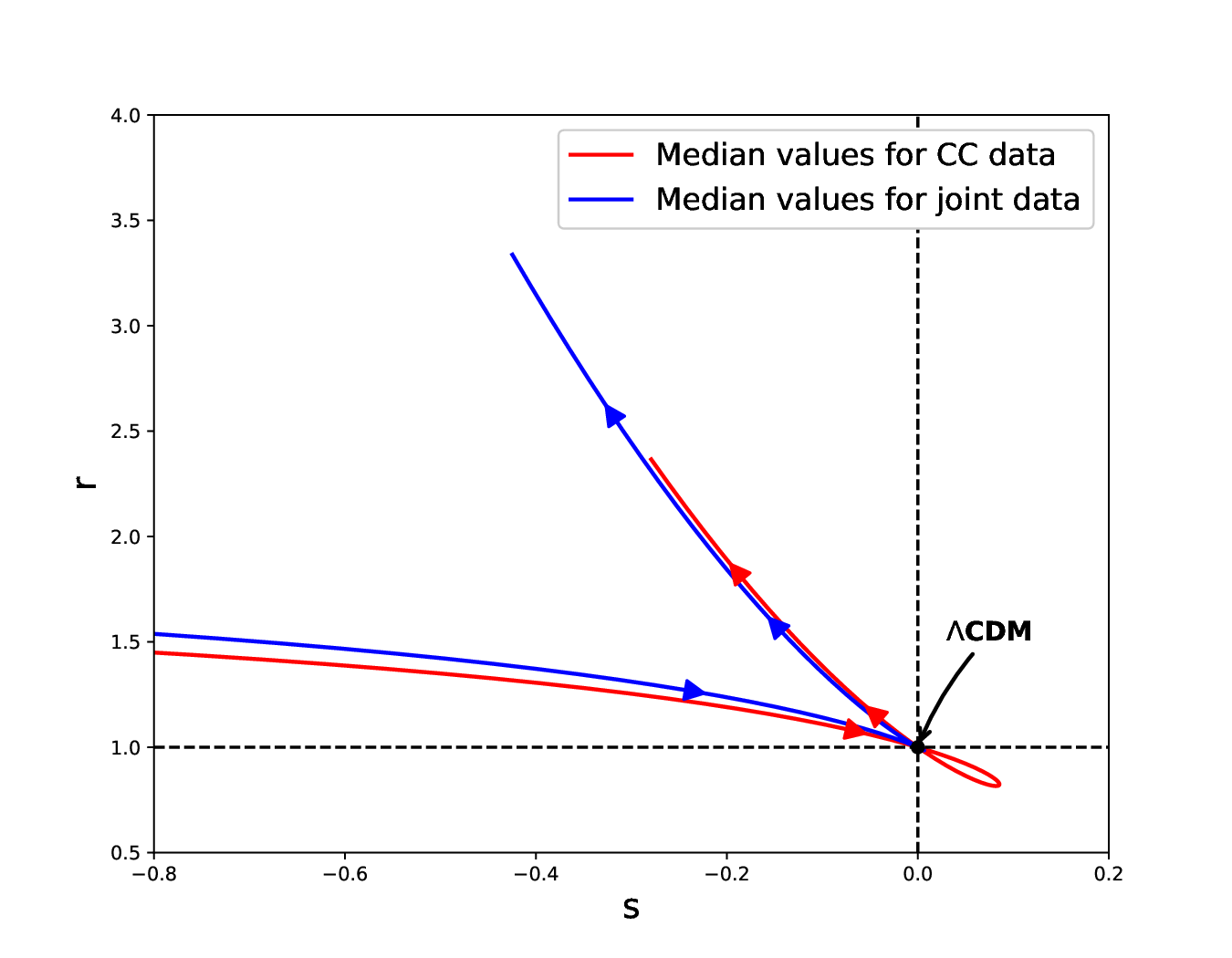}
	\caption{\textbf{For Model-1:} Plot of $s$ and $r$ plane.}
	\label{fig:15}
\end{figure}
%%%%%%%%%%%%%%%%%%%%%%%%%%%%%%%%%%%%%%%%%%%%%%%%%%%%%%%%%%%
\subsection{Statefinder diagnostic analysis}\label{sec:6.3}
\begin{figure}[ht]
	\centering
	\includegraphics[width=11cm, height=6cm]{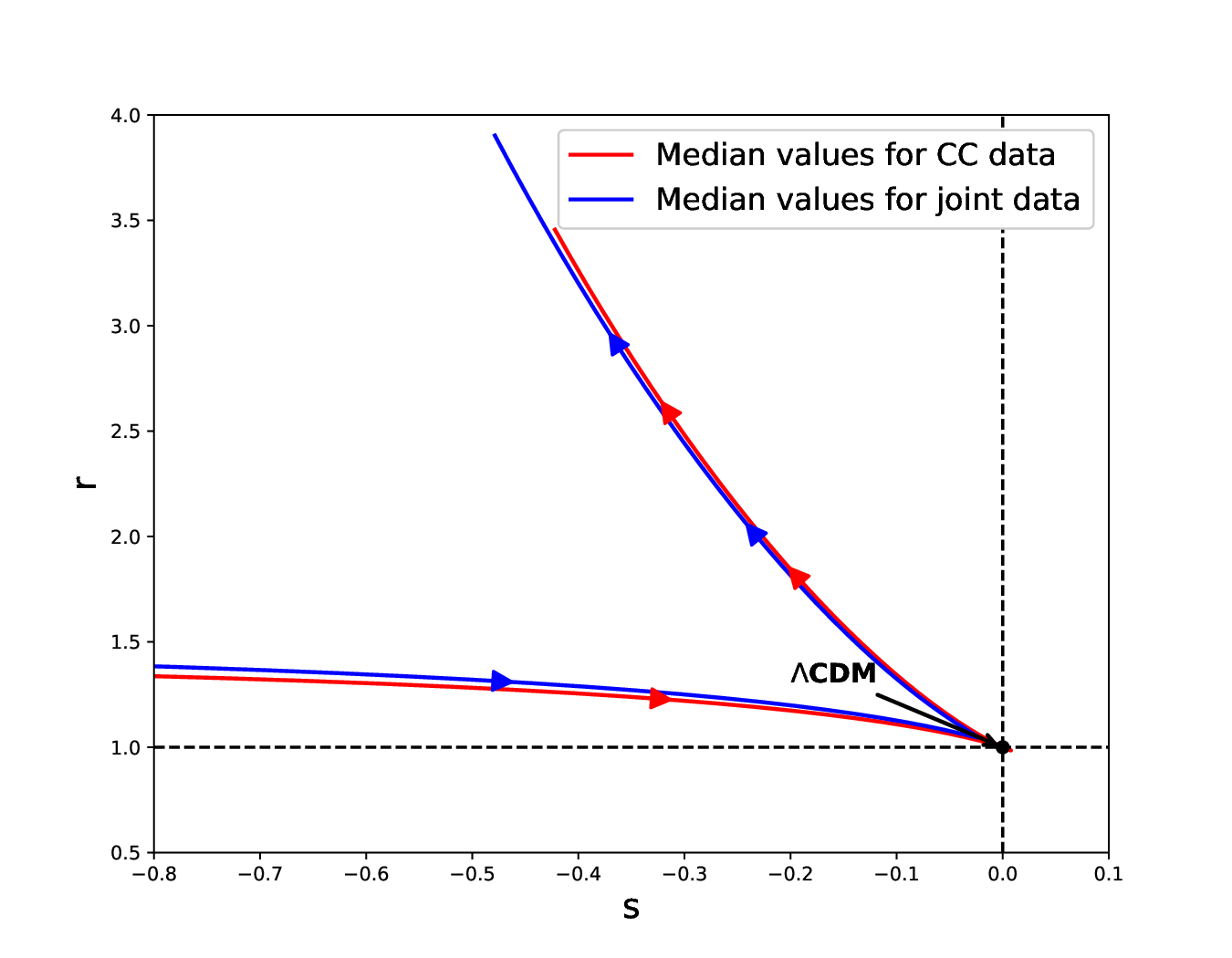}
	\caption{\textbf{For Model-2:} Plot of $s$ and $r$ plane.}
	\label{fig:16}
\end{figure}

The role of geometric parameters in characterizing the cosmological evolution of a model is well established. However, to rigorously investigate alternative DE scenarios that deviate from the standard $\Lambda$CDM framework, it becomes necessary to extend beyond the conventional descriptors such as the Hubble parameter ($H$) and the deceleration parameter ($q$). In this regard, higher-order derivatives of the scale factor $a(t)$ provide a more sensitive and refined diagnostic of cosmic dynamics, capable of distinguishing subtle differences among competing DE models. The statefinder diagnostic, introduced through the pair  $\left\{r, s\right\}$~\cite{Sahni2003}, serves as a powerful geometrical tool for this purpose. It enables a systematic classification of DE models based on their evolutionary trajectories in the  $\left\{r, s\right\}$ parameter space. The statefinder parameters  $\left\{r, s\right\}$ are defined as follows
\begin{equation}{\label{37}}
r=\frac{\dddot a}{aH^{3}}= q+2q^{2}+(1+z)\frac{dq}{dz},
\end{equation}
\begin{equation}{\label{38}}
 s= \frac{r-1}{3(q-\frac{1}{2})}, \quad \text{where} \quad q\neq \frac{1}{2}.
\end{equation}
The evolutionary behavior of various DE models in the literature is effectively characterized by their distinct values of the statefinder pair $\left\{r, s\right\}$:
\begin{itemize}
	\item In the Chaplygin gas (CG) model, the parameters typically satisfy ($r >1$, $s <0$).
	
	\item The standard $\Lambda$CDM model is characterized by $r = 1$ and $s = 0$.
	
	\item For Quintessence model, one generally finds ($r <1$, $s >0$).
	
	\item Holographic dark energy (HDE) model, ($r=1$, $s = \frac{2}{3}$).
	
	\item For Standard cold dark matter (SCDM), one may have ($r=1$, $s=1$).
	
\end{itemize}
The statefinder trajectories in the $\left\{r, s\right\}$ plane for Model-1 and Model-2 are illustrated in Figures~(\ref{fig:15}) and (\ref{fig:16}), respectively. In both models, the trajectories originate in the Chaplygin gas regime at early cosmic times, subsequently evolve through the $\Lambda$CDM fixed point and ultimately approach a unified dark sector behavior at late times. This systematic evolution underscores the ability of the proposed models to effectively interpolate between distinct cosmological phases, thereby capturing the transition from an early-time Chaplygin gas-dominated regime to a late-time unified dark matter-dark energy scenario. Such dynamical features of the statefinder trajectories are in good agreement with results reported in previous studies~\cite{fei2013statefinder}.
%%%%%%%%%%%%%%%%%%%%%%%%%%%%%%%%%%%%%%%%%%%%%%%%%%%%%%%%%%%
%%%%%%%%%%%%%%%%%%%%%%%%%%%%%%%%%%%%%%%%%%%%%%%%%%%%%%%%%%%
%%%%%%%%%%%%%%%%%%%%%%%%%%%%%%%%%%%%%%%%%%%%%%%%%%%%%%%%%%%
%%%%%%%%%%%%%%%%%%%%%%%%%%%%%%%%%%%%%%%%%%%%%%%%%%%%%%%%%%
\subsection{Om($\mathit{z}$) diagnostics analysis}\label{sec:6.4}
\begin{figure}[ht]
	\centering
	\includegraphics[width=13cm, height=6cm]{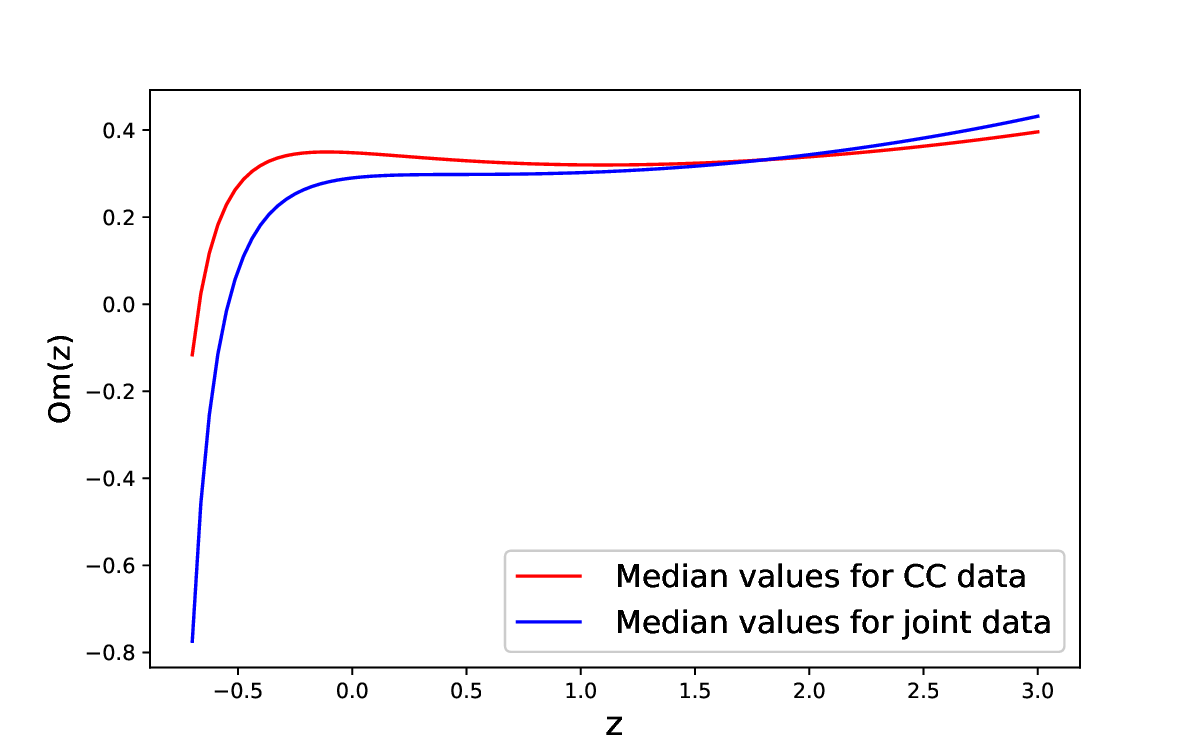}
	\caption{\textbf{For Model-1:} Plot of Om($\mathit{z}$) diagnostics with $z$.}
	\label{fig:17}
\end{figure}

\begin{figure}[ht]
	\centering
	\includegraphics[width=13cm, height=6cm]{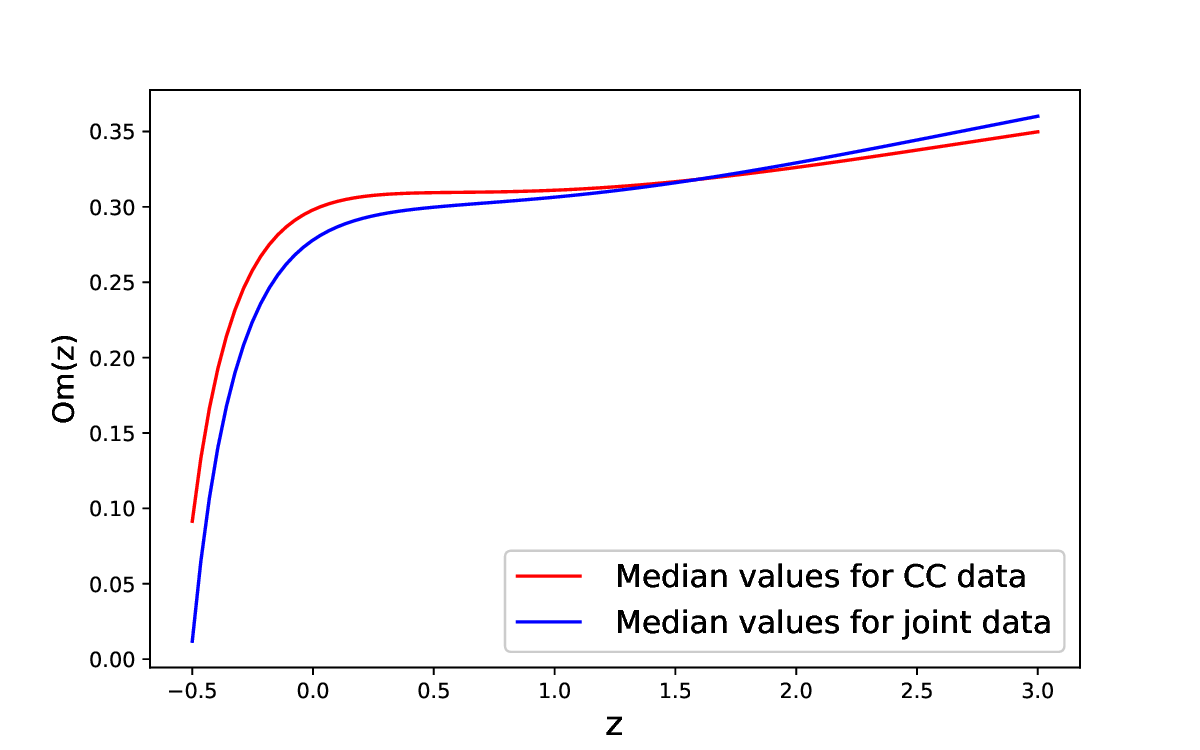}
	\caption{\textbf{For Model-2:} Plot of Om($\mathit{z}$) diagnostics with $z$.}
	\label{fig:18}
\end{figure}

The Om($z$) diagnostic constitutes a powerful geometrical probe for distinguishing between different dark energy models~\cite{sahni2008two}. For a spatially flat universe, it is given by
\begin{equation} {\label{39}}
	Om(z) = \frac{\left(\frac{H(z)}{H_{0}}\right)^{2}-1}{(1+z)^{3}-1}.
\end{equation}
The slope of the Om($z$) diagnostic reflects the nature of dark energy: a positive slope indicates phantom behavior, while a negative slope corresponds to quintessence. A constant Om($z$) is characteristic of the $\Lambda$CDM scenario. The evolution of the Om($z$) diagnostic for the considered parametrizations, obtained using CC and joint datasets, is shown in Figs.~(\ref{fig:17}) and (\ref{fig:18}). The deviation of Om($z$) from a constant behavior indicates a departure from the standard $\Lambda$CDM cosmology and supports the presence of dynamical dark energy. At the present epoch ($z \approx 0$), the values of Om($z$) remain within the observationally acceptable range, ensuring consistency with current cosmological constraints. As the redshift increases, Om($z$) exhibits a smooth variation and tends to converge at higher redshift, reflecting the expected matter-dominated behavior in the early universe. In the negative redshift regime ($z < 0$), the behavior of Om($z$) reflects the continued evolution of the expansion dynamics of the models, which may be associated with a super-accelerating phase depending on their underlying behavior.
%%%%%%%%%%%%%%%%%%%%%%%%%%%%%%%%%%%%%%%%%%%%%%%%%%%%%%%%%%%
%%%%%%%%%%%%%%%%%%%%%%%%%%%%%%%%%%%%%%%%%%%%%%%%%%%%%%%%%%%
%%%%%%%%%%%%%%%%%%%%%%%%%%%%%%%%%%%%%%%%%%%%%%%%%%%%%%%%%%%
%%%%%%%%%%%%%%%%%%%%%%%%%%%%%%%%%%%%%%%%%%%%%%%%%%%%%%%%%%%
%%%%%%%%%%%%%%%%%%%%%%%%%%%%%%%%%%%%%%%%%%%%%%%%%%%%%%%%%%%
\begin{figure}[!htb]
	\captionsetup{skip=0.4\baselineskip,size=footnotesize}
	\begin{minipage}{0.50\textwidth}
		\centering
		\includegraphics[width=8.6cm,height=7cm]{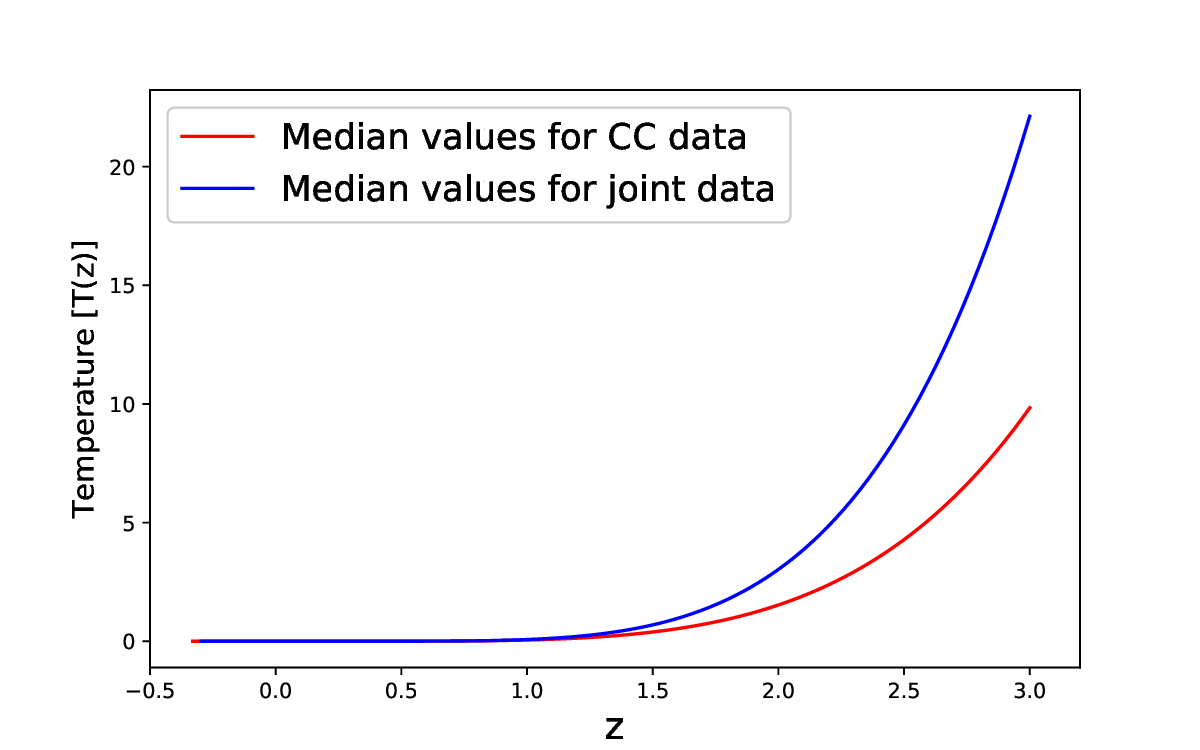}
		\caption{\textbf{For Model-1:} Plot of temperature with $\mathit{z}$.}
		\label{fig:19}
	\end{minipage}\hfill
	\begin{minipage}{0.50\textwidth}
		\centering
		\includegraphics[width=8.6cm,height=7cm]{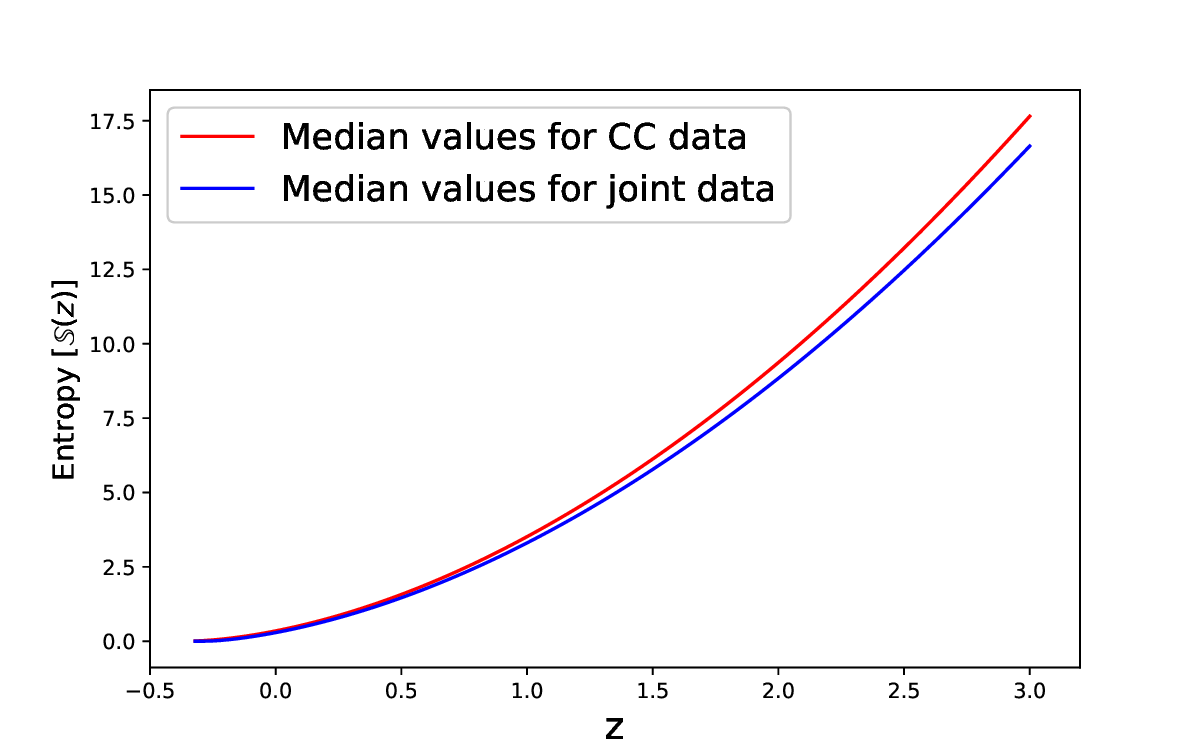}
		\caption{\textbf{For Model-1:} Plot of entropy density with $\mathit{z}$.}
		\label{fig:20}
	\end{minipage}
\end{figure}
%%%%%%%%%%%%%%%%%%%%%%%%%%%%%%%%%%%%%%%%%%%%%%%%%%%%%%%%%%%%%%%
%%%%%%%%%%%%%%%%%%%%%%%%%%%%%%%%%%%%%%%%%%%%%%%%%%%%%%%%%%%
\begin{figure}[!htb]
	\captionsetup{skip=0.4\baselineskip,size=footnotesize}
	\begin{minipage}{0.50\textwidth}
		\centering
		\includegraphics[width=8.6cm,height=7cm]{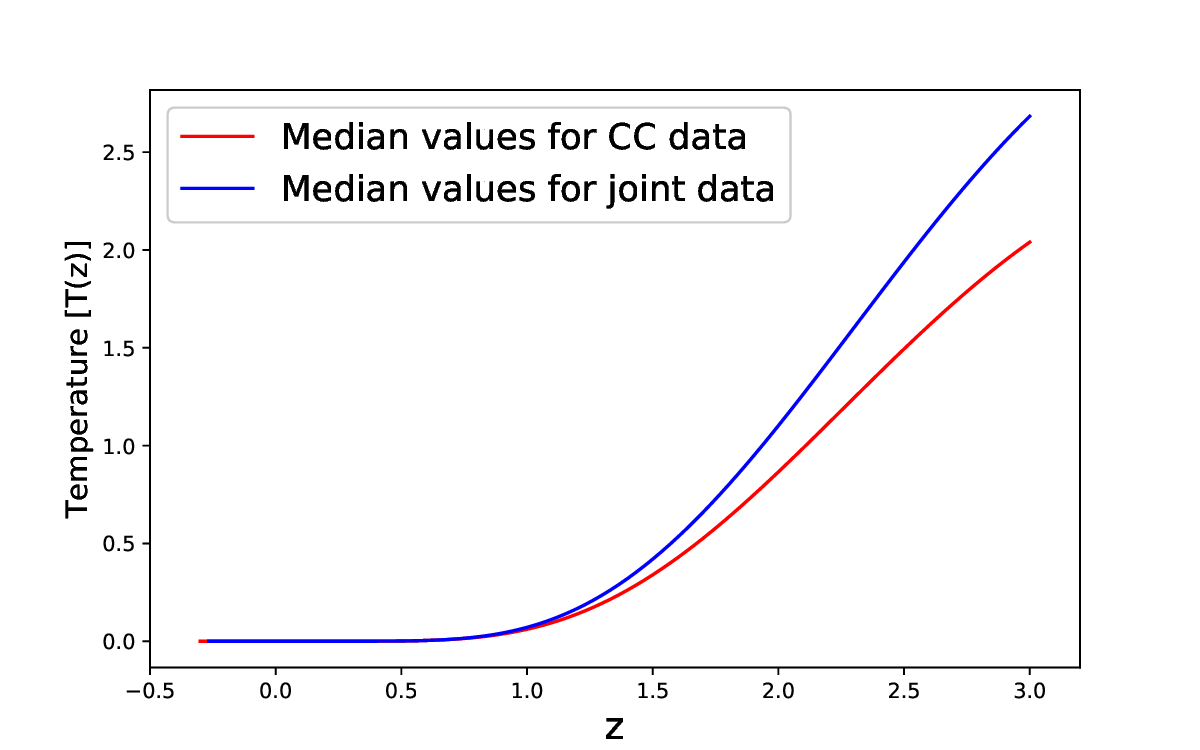}
		\caption{\textbf{For Model-2:} Plot of temperature with $\mathit{z}$.}
		\label{fig:21}
	\end{minipage}\hfill
	\begin{minipage}{0.50\textwidth}
		\centering
		\includegraphics[width=8.6cm,height=7cm]{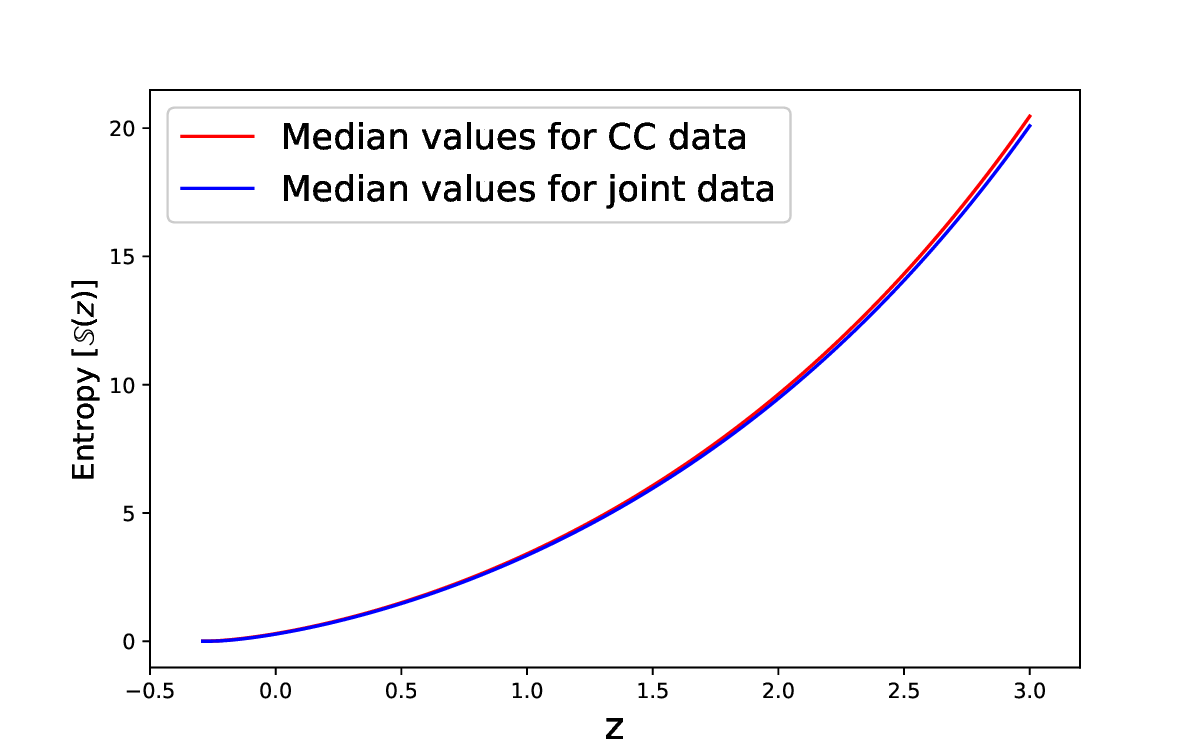}
		\caption{\textbf{For Model-2:} Plot of entropy density with $\mathit{z}$.}
		\label{fig:22}
	\end{minipage}
\end{figure}
%%%%%%%%%%%%%%%%%%%%%%%%%%%%%%%%%%%%%%%%%%%%%%%%%%%%%%%%%%%%%%%
%%%%%%%%%%%%%%%%%%%%%%%%%%%%%%%%%%%%%%%%%%%%%%%%%%%%%%%%%%%%%%%%%%%%%
\subsection{Study of thermodynamic viability and entropy dynamics in an evolving universe}\label{sec:6.5}
The thermodynamic characteristics of the universe are explored in the setting of $f(T)$ gravity by considering the first law of thermodynamics for an ideal fluid within a finite volume $V$~\cite{salako2013lambdacdm,bamba2012reconstruction}.
\begin{equation} {\label{41}}
T ds = d(V \rho)+p dV,
\end{equation}
we can rewrite it as follows:
\begin{equation} {\label{42}}
	T ds = d((\rho+p)V)- V dp
\end{equation}
combined with the use of the following thermodynamic relation:
\begin{equation} {\label{43}}
	dp = \left[\frac{\rho+p}{T}\right] dT.
\end{equation}
From this, the differential expression for the entropy ($s$) is obtained as
\begin{equation} {\label{44}}
ds = \frac{ d(V(\rho+p))}{T}- V(\rho+p) \frac{dT}{T^{2}},
\end{equation}
which can be simplified to:
\begin{equation} {\label{45}}
	ds = d\left(\frac{ V(\rho+p)}{T}\right).
\end{equation}
Upon performing the integration, the total entropy takes the form
\begin{equation} {\label{46}}
	s = \frac{ V(\rho+p)}{T}.
\end{equation}
The quantity ($\mathbb{S}$), representing the entropy density, is defined by
\begin{equation} {\label{47}}
\mathbb{S}=\frac{s}{V}= \frac{ (\rho+p)}{T}= \frac{ \rho(\omega+1)}{T}.
\end{equation}
Assuming the cosmic fluid obeys the barotropic relation $p=\omega \rho$, with ($0<\omega<1$), the corresponding form of the first law of thermodynamics becomes:
\begin{equation} {\label{48}}
d(V \rho)+\omega \rho dV= (1+\omega)T d\left(\frac{V\rho}{T}\right).
\end{equation}
On integrating the differential form:
\begin{equation} {\label{49}}
\omega d\rho = (1+\omega)\rho \frac{dT}{T},
\end{equation}
from which the temperature can be expressed in terms of the energy density:
\begin{equation} {\label{50}}
T=\rho^{\left(\frac{\omega}{\omega+1}\right)};
\end{equation}
for Model-1:
\begin{equation} {\label{51}}
	T=\left[\frac{(2n-1)}{2} \alpha 6^{n} H_{0}^{2n} (1+z)^{2n(1+q_{0})} \exp\left[n q_{1}\left(\log[1+z]\right)^{2}\right]\right]^{\left\{\frac{\left[-1+\frac{2n}{3}\left(1+q_{0}+q_{1} \log(1+z)\right)\right]}{\frac{2n}{3}\left(1+q_{0}+q_{1} \log(1+z)\right)}\right\}}, \qquad \qquad 
\end{equation}
and for Model-2:
\begin{equation} {\label{52}}
T=\left[\frac{(2n-1)}{2} \alpha 6^{n} H_{0}^{2n} (1+z)^{2n(1+q_{0})} \exp\left[2n q_{1}\left(1-\cos\left(\log[1+z]\right)\right)\right]\right]^{\left\{\frac{\left[-1+\frac{2n}{3}\left[1+q_{0}+q_{1} \sin\left(\log(1+z)\right)\right]\right]}{\frac{2n}{3}\left[1+q_{0}+q_{1} \sin\left(\log(1+z)\right)\right]}\right\}}.
\end{equation}
Substituting this result back into the expression, we obtain the entropy density as:
\begin{equation} {\label{53}}
	\mathbb{S}=(\omega+1)\rho^{\left(\frac{1}{\omega+1}\right)};
\end{equation}
for Model-1:
\begin{equation} {\label{54}}
\mathbb{S}= \left(\frac{2n}{3}\left(1+q_{0}+q_{1} \log(1+z)\right)\right)\left[\frac{(2n-1)}{2} \alpha 6^{n} H_{0}^{2n} (1+z)^{2n(1+q_{0})} \exp\left[n q_{1}\left(\log[1+z]\right)^{2}\right]\right]^{\left\{\frac{1}{\frac{2n}{3}\left(1+q_{0}+q_{1} \log(1+z)\right)}\right\}},
\end{equation} 
and for Model-2:
\begin{equation} {\label{55}}
\resizebox{1.0\textwidth}{!}{$	
\mathbb{S}= \left(\frac{2n}{3}\left[1+q_{0}+q_{1} \sin\left(\log(1+z)\right)\right]\right)\left[\frac{(2n-1)}{2} \alpha 6^{n} H_{0}^{2n} (1+z)^{2n(1+q_{0})} \exp\left[2n q_{1}\left(1-\cos\left(\log[1+z]\right)\right)\right]\right]^{\left\{\frac{1}{\frac{2n}{3}\left[1+q_{0}+q_{1} \sin\left(\log(1+z)\right)\right]}\right\}}.
$}
\end{equation} 
\vspace{0.2cm}\\
The temperature and entropy density evolution of the considered cosmological models, presented in Figs.~(\ref{fig:19}) to (\ref{fig:22}), clearly illustrates their dependence on redshift under observational constraints. From Figs.~(\ref{fig:19}) and (\ref{fig:21}), it is observed that the universe possessed a much higher temperature in the early stages of its evolution, with temperature increasing gradually as redshift increases. The thermal properties associated with the models exhibit consistency with conventional cosmological frameworks, thereby supporting the model’s capability to capture the expected thermal history of the universe. In the far-future regime $(z<0)$, the temperature approaches a constant value asymptotically, reflecting the tendency of the system to evolve toward thermal equilibrium in the low-energy regime. Furthermore, the entropy density evolution, depicted in Figs.~(\ref{fig:20}) and (\ref{fig:22}), is obtained using the dimensionless energy density normalized with respect to the critical density scale $3H_0^2$. The results show that entropy density increases with redshift, implying that the early universe was characterized by a significantly higher entropy content. This trend is in agreement with the standard expectation that both energy density and temperature were larger in the early cosmic phases. As the universe expands, the entropy density decreases gradually due to volume expansion and energy dilution; whereas the total entropy persistently grows, adhering to the generalized second law. Moreover, the derived relation $\mathbb{S} \propto \rho^{\left(\frac{1}{\omega+1}\right)}$ demonstrates a significant connection between thermodynamic evolution and the dynamics of the energy density. These outcomes verify the internal coherence of the thermodynamic treatment and further support the physical acceptability of the proposed $f(T)$ gravity formulation as a robust framework for explaining the evolutionary history of the universe. 
%%%%%%%%%%%%%%%%%%%%%%%%%%%%%%%%%%%%%%%%%%%%%%%%%%%%%%%%%%%
\subsection{Estimation of the age of the universe}\label{sec:6.6}
The evolution of the cosmic age $t(z)$, expressed as a function of redshift $z$ for the cosmological model is given by~\cite{tong2009cosmic}
\begin{equation} {\label{40}}
	t(z) = \int_{z}^{\infty} \frac{dz}{(1+z)H(z)}.
\end{equation}
The current age of the universe ($t_{0}$), is determined by numerically evaluating the integral of the Hubble parameter $H(z)$ (from Eqs.~(\ref{22}) and (\ref{24})) at the present epoch ($z = 0$). For Model-1, we obtain $t_{0} = 12.73$ Gyr (CC dataset) and $t_{0} = 12.52$ Gyr (joint dataset). For Model-2, the present age is found to be $t_{0} = 14.61$ Gyr and $t_{0} = 14.82$ Gyr for the CC and joint datasets, respectively. These estimates are in agreement with recent observational studies and support the viability of the proposed models in describing the late-time expansion of the universe~\cite{tatum2025extracting,Valcin2020}.
%%%%%%%%%%%%%%%%%%%%%%%%%%%%%%%%%%%%%%%%%%%%%%%%%%%%%%%%%%%
%%%%%%%%%%%%%%%%%%%%%%%%%%%%%%%%%%%%%%%%%%%%%%%%%%%%%%%%%%%%%%%%%
\section{Conclusions}\label{sec:7}
In the present study, we investigate the cosmological evolution of the universe within the framework of $f(T)$ gravity in a spatially flat FLRW background by adopting a power-law form of the function $f(T)=\alpha(-T)^{n}$, together with two distinct parametrizations of the deceleration parameter, namely the logarithmic form $q(z)=q_{0}+q_{1}\log(1+z)$ and the log-periodic form $q(z)=q_{0}+q_{1}\sin[\log(1+z)]$. These parametrizations provide a flexible and dynamically rich description of the expansion history, allowing both smooth and oscillatory features in cosmic evolution. The model parameters $H_{0}$, $q_{0}$ and $q_{1}$ have been constrained using the cosmic chronometer (CC) dataset as well as the joint (CC+Pantheon) dataset through a Bayesian MCMC analysis and the corresponding results are presented in Tables~(\ref{table:1}) and (\ref{table:2}). The best-fit Hubble parameter $H(z)$ curve, shown in Figs.~(\ref{fig:3}) and (\ref{fig:4}), indicates that both models are in good agreement with the measured cosmic chronometer observations, thereby demonstrating the reliability of the proposed models.
\vspace{0.2cm}\\
The evolution of the deceleration parameter (Figs.~(\ref{fig:1}) and (\ref{fig:2})), clearly demonstrates that both models successfully describe the transition from an early decelerating phase to the present accelerated expansion. The present-day values are obtained as $q_{0}=-0.478$ (CC) and $q_{0}=-0.565$ (joint) for Model-1, while for Model-2 they are $q_{0}=-0.553$ (CC) and $q_{0}=-0.583$ (joint), confirming the current accelerated state of the universe. The transition redshift is found to be $z_{t}=0.6692$ for Model-1 and $z_{t}=0.6563$ for Model-2, which are consistent with observational expectations.
%\vspace{0.2cm}\\
An important feature emerging from the present analysis is that both parametrizations allow for a transition into a super-accelerated (phantom) regime at late times. In particular, the log-periodic parametrization provides an alternative functional description of the expansion dynamics compared to the logarithmic case, thereby offering a broader perspective on cosmic evolution. This highlights the significance of generalized parametrizations in describing the evolving nature of DE.
\vspace{0.2cm}\\
The physical viability of the models is further supported by the evolution of the dark energy density and pressure, as shown in Figs.~(\ref{fig:7}) to (\ref{fig:10}). The energy density remains positive throughout the cosmic evolution, ensuring physical consistency, while the pressure becomes negative at late times, thereby driving the accelerated expansion. The corresponding behavior of the EoS parameter, depicted in Figs.~(\ref{fig:11}) and (\ref{fig:12}), indicates that the universe currently lies in the quintessence regime, followed by a smooth transition into the phantom domain at late times, suggesting a quintom-like nature of DE. At higher redshifts, both models approach a matter-dominated phase, consistent with the requirements of structure formation.
\vspace{0.2cm}\\
The analysis of energy conditions (see Figs.~(\ref{fig:13}) and (\ref{fig:14})) reveals that the Null, Weak and Dominant Energy Conditions are satisfied up to the present epoch, while the Strong Energy Condition is violated, which is necessary for explaining the observed accelerated expansion. At late times, the violation of the NEC further indicates the emergence of phantom behavior. Additional support is provided by geometrical diagnostics such as the statefinder $\{r,s\}$ plane and the $Om(z)$ diagnostic, both of which indicate a clear deviation from the standard $\Lambda$CDM model and favor a dynamically evolving DE scenario.
\vspace{0.2cm}\\
The thermal characteristics of the model are examined by studying the evolution of temperature and entropy density. The temperature $T(z)$ increases with redshift, indicating a hot early universe followed by gradual cooling as expansion proceeds. Similarly, entropy density $\mathbb{S}(z)$ is higher at large redshift, reflecting the greater energy content in the early universe and decreases progressively at lower redshift due to cosmic expansion. Overall, this behavior indicates a physically consistent thermodynamic evolution within the framework of the study. The estimated age of the universe is found to be $t_{0}=12.73$ Gyr and $t_{0}=12.52$ Gyr for Model-1, and $t_{0}=14.61$ Gyr and $t_{0}=14.82$ Gyr for Model-2, corresponding to the CC and joint datasets, respectively. These values are in line with recent observational estimates, thereby reinforcing the viability of this framework.
\vspace{0.2cm}\\
To summarize, this study explores the cosmic evolution of the universe within modified teleparallel gravity using a power-law form of $f(T)$ along with logarithmic and log-periodic parametrizations of the deceleration parameter. Both models successfully reproduce the transition from an early decelerated phase to the present accelerated expansion and describe its subsequent dynamical evolution in a physically coherent manner. The results show that the proposed framework effectively captures the essential features of late-time cosmic dynamics in a simple and well-behaved formulation. Overall, the models provide a reliable and phenomenologically useful description of DE evolution, offering valuable insight into the late-time dynamics of the universe within the framework of $f(T)$ gravity.
%%%%%%%%%%%%%%%%%%%%%%%%%%%%%%%%%%%%%%%%%%%%%
\section*{\textbf{Acknowledgements}}
GPS is thankful to the Inter-University Centre for Astronomy and Astrophysics (IUCAA), Pune, India for support under Visiting Associateship program.
%%%%%%%%%%%%%%%%%%%%%%%%%%%%%%%%%%%%%%%%%%%%%%%%%%
%\section*{\textbf{Data Availability statement}}
%No data was used for the research described in the article.
%%%%%%%%%%%%%%%%%%%%%%%%%%%%%%%%%%%%%%%%%%%%%%%%%
%\section*{\textbf{Declaration of competing interest}}
%The authors declare that they have no known competing financial interests or personal relationships that could have appeared to influence the work reported in this paper. 
%%%%%%%%%%%%%%%%%%%%%%%%%%%%%%%%%%%%%%%%%%%%%%%%%%%%%%%%%%%%%%%%%%


\begin{thebibliography}{0}
	%1	
	\bibitem{1998AJ....116.1009R} A. G. Riess, A. V.  Filippenko, P. Challis, A. Clocchiatti, A. Diercks  et al., Observational Evidence from Supernovae for an Accelerating Universe and a Cosmological Constant, Astronomical Journal, \textbf{116}, 1009-1038 (1998) \url{https://doi.org/10.1086/300499} 
	%2
	\bibitem{1999ApJ...517..565P}  S. Perlmutter, G. Aldering, G. Goldhaber, R. A. Knop, P. Nugent et al., Measurements of $\Omega$ and $\Lambda$ from 42 High-Redshift Supernovae, Astrophysical Journal, \textbf{517}, 565–586 (1999)  \url{https://doi.org/10.1086/307221} 
	%3
	\bibitem{2020A&A...641A...6P} N. Aghanim, Y. Akrami, M. Ashdown, J. Aumont, C. Baccigalupi, et al., Planck 2018 results. VI. Cosmological parameters, Astronomy and Astrophysics, \textbf{641}, A6 (2020) \url{https://doi.org/10.1051/0004-6361/201833910}
	%4
	\bibitem{weinberg1989cosmological} S. Weinberg, The cosmological constant problem, Reviews of Modern Physics, \textbf{61}, 1 (1989) \url{https://doi.org/10.1103/RevModPhys.61.1}
	%5
	\bibitem{di2021realm} E. Di Valentino, O. Mena, S. Pan, L. Visinelli, W. Yang, et al., In the realm of the Hubble tension—a review of solutions, Classical and Quantum Gravity, \textbf{38}, 153001 (2021) \url{https://doi.org/10.1088/1361-6382/ac086d}
	%6
	\bibitem{carroll2001cosmological} S. M. Carroll, The cosmological constant, Living Reviews in Relativity, \textbf{4}, 1-56 (2001) \url{https://doi.org/10.12942/lrr-2001-1}
	%6
	\bibitem{padmanabhan2003cosmological} T. Padmanabhan, Cosmological constant—the weight of the vacuum, Physics reports, \textbf{380}, 235--320 (2003) \url{https://doi.org/10.1016/S0370-1573(03)00120-0}
	%6
	\bibitem{copeland2006dynamics} E. J. Copeland, M. Sami, S. Tsujikawa, Dynamics of dark energy, International Journal of Modern Physics D, \textbf{15}, 1753--1935 (2006) \url{https://doi.org/10.1142/S021827180600942X}
	%7
	\bibitem{buchdahl1970non} H. A. Buchdahl, Non-linear Lagrangians and cosmological theory, Monthly Notices of the Royal Astronomical Society, \textbf{150}, 1-8 (1970) \url{https://doi.org/10.1093/mnras/150.1.1}
	%8
	\bibitem{harko2011f} T. Harko, F. S. N. Lobo, S. Nojiri, S. D. Odintsov, $f(R,T)$ gravity, Physical Review D, \textbf{84}, 024020 (2011) \url{https://doi.org/10.1103/PhysRevD.84.024020}
	%9
	\bibitem{nojiri2011unified} S. Nojiri, S. D. Odintsov, Unified cosmic history in modified gravity: from $f(R)$ theory to Lorentz non-invariant models, Physics Reports, \textbf{505}, 59-144 (2011) \url{https://doi.org/10.1016/j.physrep.2011.04.001}
	%9
	\bibitem{jimenez2018coincident} J. B. Jim{\'e}nez, L. Heisenberg, T. Koivisto, Coincident general relativity, Physical Review D, \textbf{98}, 044048 (2018) \url{https://doi.org/10.1103/PhysRevD.98.044048}
	%10
	\bibitem{nojiri2017modified} S. Nojiri, S. D. Odintsov, V. K. Oikonomou, Modified gravity theories on a nutshell: Inflation, bounce and late-time evolution, Physics Reports, \textbf{692}, 1-104 (2017) \url{https://doi.org/10.1016/j.physrep.2017.06.001}
	%10
	\bibitem{bamba2010finite} K. Bamba, S. D. Odintsov, L. Sebastiani, S. Zerbini, Finite-time future singularities in modified Gauss--Bonnet and  f(R, G) gravity and singularity avoidance, The European Physical Journal C, \textbf{67}, 295--310 (2010) \url{https://doi.org/10.1140/epjc/s10052-010-1292-8}
	%10
	\bibitem{elizalde2010lambdacdm} E. Elizalde,and R. Myrzakulov, V. V. Obukhov, D. S{\'a}ez-G{\'o}mez, $\Lambda$CDM epoch reconstruction from F (R, G) and modified Gauss--Bonnet gravities, Classical and Quantum Gravity, \textbf{27}, 095007 (2010) \url{https://doi.org/10.1088/0264-9381/27/9/095007}
	%25
	\bibitem{harko2010f} T. Harko, F. S. N. Lobo, $f(R, L_m)$ gravity, The European Physical Journal C, \textbf{70}, 373--379 (2010) \url{https://doi.org/10.1140/epjc/s10052-010-1467-3}
	%12
	\bibitem{capozziello2019extended} S. Capozziello, R. D'Agostino, O. Luongo, Extended gravity cosmography, International Journal of Modern Physics D, \textbf{28}, 1930016 (2019) \url{https://doi.org/10.1142/S0218271819300167}
	%11
	\bibitem{capozziello2023role} S. Capozziello, V. De Falco, C. Ferrara, The role of the boundary term in $f(Q, B)$ symmetric teleparallel gravity,The European Physical Journal C, \textbf{83}, 915 (2023) \url{https://doi.org/10.1140/epjc/s10052-023-12072-y}
	%14
	\bibitem{kotambkar2017anisotropic} S. Kotambkar, G. P. Singh, R. Kelkar, B. K. Bishi, Anisotropic Bianchi type I cosmological models with generalized Chaplygin gas and dynamical gravitational and cosmological constants, Communications in Theoretical Physics, \textbf{67}, 222 (2017) \url{https://doi.org/10.1088/0253-6102/67/2/222}
	%15
	%\bibitem{singh1997new} G. P. Singh, K. Desikan, A new class of cosmological models in Lyra geometry, Pramana, \textbf{49}, 205--212 (1997) \url{https://doi.org/10.1007/BF02845856}
	%13
	\bibitem{lalke2023late} A. R. Lalke, G. P. Singh, A. Singh, Late-time acceleration from ekpyrotic bounce in $f(Q, T)$ gravity, International Journal of Geometric Methods in Modern Physics, \textbf{20}, 2350131 (2023) \url{doi:https://doi.org/10.1142/S0219887823501311}
	%19
	\bibitem{hulke2020variable} N. Hulke, G. P. Singh, B. K. Bishi, A. Singh, Variable Chaplygin gas cosmologies in $f(R, T)$ gravity with particle creation, New Astronomy, \textbf{77}, 101357 (2020) \url{https://doi.org/10.1016/j.newast.2020.101357}
	%20
	%\bibitem{garg2024cosmological} R. Garg, G. P. Singh, A. R. Lalke, S. Ray, Cosmological model with linear equation of state parameter in $f(R, L_m)$ gravity, Physics Letters A, \textbf{525}, 129937 (2024) \url{https://doi.org/10.1016/j.physleta.2024.129937}
	%16
	\bibitem{singh2025observational} A. Singh, S. Mandal, R. Chaubey, R. Raushan, Observational constraints on the expansion scalar and shear relation in the Locally rotationally symmetric Bianchi I model, Physics of the Dark Universe, \textbf{47}, 101798 (2025) \url{https://doi.org/10.1016/j.dark.2024.101798}
	%22
	\bibitem{singh2024conservative} K. N. Singh, G. R. P. Teruel, S. K. Maurya, T. Chowdhury, F. Rahaman, Conservative wormholes in generalized $K(R, T)$ function, Journal of High Energy Astrophysics, \textbf{44}, 132--145 (2024) \url{https://doi.org/10.1016/j.jheap.2024.09.009}
	%17
	%\bibitem{varela2025cosmological} M. B. Varela, O. Bertolami, Is cosmological data suggesting a nonminimal coupling between matter and gravity?, Physics of the Dark Universe, \textbf{48}, 101861 (2025) \url{https://doi.org/10.1016/j.dark.2025.101861}
	%18
	%\bibitem{singh2022cosmic} A. Singh, G. P. Singh, A. Pradhan, Cosmic dynamics and qualitative study of Rastall model with spatial curvature, International Journal of Modern Physics A, \textbf{37}, 2250104 (2022) \url{https://doi.org/10.1142/S0217751X22501044}
	%21
	\bibitem{chaudhary2025extracting} H. Chaudhary, S. K. J. Pacif, G. Mustafa, F. Atamurotov, F. Javed, Extracting H0 and rd in $q(t)$ parametrization models, Journal of High Energy Astrophysics, \textbf{45}, 340--349 (2025) \url{https://doi.org/10.1016/j.jheap.2025.01.001}
	%23
	\bibitem{goswami2024flrw} G. K. Goswami, R. Rani, J. K. Singh, A. Pradhan, FLRW cosmology in Weyl type $f(Q)$ gravity and observational constraints, Journal of High Energy Astrophysics, \textbf{43}, 105--113 (2024) \url{https://doi.org/10.1016/j.jheap.2024.06.011}
	%24
	\bibitem{shukla2025multi} B. K. Shukla, S. Sahlu, D. Sofuo{\u{g}}lu, P. Mishra, A. H. Alfedeel, Multi-components fluid in $f(R, T)$ gravity with observational constraints, The European Physical Journal Plus, \textbf{140}, 1--14 (2025)
	\url{https://doi.org/10.1140/epjp/s13360-025-06200-8}
	%24
	\bibitem{escamilla2024exploring} L. A. Escamilla, D. Fiorucci, G. Montani, E. Di Valentino, Exploring the Hubble tension with a late time Modified Gravity scenario, Physics of the Dark Universe, \textbf{46}, 101652 (2024)
	\url{https://doi.org/10.1016/j.dark.2024.101652}
	%16
	\bibitem{patle2026dynamical} K. R. Patle, G. P. Singh, R. Garg, Dynamical constraints on variable vacuum energy in Brans-Dicke theory, arXiv preprint arXiv:2601.00419, (2026) \url{https://doi.org/10.48550/arXiv.2601.00419}
	%28
	\bibitem{Bengochea} G. R. Bengochea, R. Ferraro, Dark torsion as the cosmic speed-up, Physical Review D, \textbf{79}, 124019 (2009)  \url{doi:https://doi.org/10.1103/PhysRevD.79.124019}
	%29
	\bibitem{cai2016f} Yi-Fu Cai, S. Capozziello, M. De Laurentis, E. N. Saridakis, f (T) teleparallel gravity and cosmology, Reports on Progress in Physics, \textbf{79}, 106901 (2016) \url{https://doi.org/10.1088/0034-4885/79/10/106901}
	%30
	\bibitem{paliathanasis2016cosmological} A. Paliathanasis, J. D. Barrow, P. G. L. Leach, Cosmological solutions of f (T) gravity, Physical Review D, \textbf{94}, 023525 (2016) \url{https://doi.org/10.1103/PhysRevD.94.023525}
	%31
	\bibitem{salako2013lambdacdm} I. G. Salako, M. E. Rodrigues, A. V. Kpadonou, M. J. S. Houndjo, J. Tossa, $\Lambda$CDM model in $f(T)$ gravity: reconstruction, thermodynamics and stability, Journal of Cosmology and Astroparticle Physics, \textbf{2013}, 060--060 (2013) \url{https://doi.org/10.1088/1475-7516/2013/11/060}
	%32
	\bibitem{capozziello2011cosmography} S. Capozziello, V. F. Cardone, H. Farajollahi, A. Ravanpak, Cosmography in $f(T)$ gravity, Physical Review D, \textbf{84}, 043527 (2011) \url{ https://doi.org/10.1103/PhysRevD.84.043527}
	%33
	\bibitem{liu2012energy} Di Liu, M. J. Reboucas, Energy conditions bounds on $f(T)$ gravity, Physical Review D, \textbf{86}, 083515 (2012) \url{ https://doi.org/10.1103/PhysRevD.86.083515}
	%34
	\bibitem{cai2011matter} Yi-Fu Cai, Shih-Hung Chen, J. B. Dent, S. Dutta, E. N. Saridakis, Matter bounce cosmology with the $f(T)$ gravity, Classical and Quantum Gravity, \textbf{28}, 215011 (2011) \url{ https://doi.org/10.1088/0264-9381/28/21/215011}
	%36
	\bibitem{zhadyranova2024exploring} A. Zhadyranova, M. Koussour, S. Bekkhozhayev, V. Zhumabekova, J. Rayimbaev, Exploring late-time cosmic acceleration: A study of a linear $f(T)$ cosmological model using observational data, Physics of the Dark Universe, \textbf{45}, 101514 (2024) \url{https://doi.org/10.1016/j.dark.2024.101514}
	%37
	\bibitem{bamba2011equation} K. Bamba, Chao-Qiang Geng, Chung-Chi Lee, Ling-Wei Luo, Equation of state for dark energy in $f(T)$ gravity, Journal of Cosmology and Astroparticle Physics, \textbf{2011}, 021--021 (2011) \url{https://doi.org/10.1088/1475-7516/2011/01/021}
	%38
	\bibitem{paliathanasis2014new} A. Paliathanasis, S. Basilakos, E. N. Saridakis, S. Capozziello, K. Atazadeh, F. Darabi, M. Tsamparlis, New Schwarzschild-like solutions in $f(T)$ gravity through Noether symmetries, Physical Review D, \textbf{89}, 104042 (2014) \url{https://doi.org/10.1103/PhysRevD.89.104042}
	%39
	\bibitem{capozziello2017model} S. Capozziello, R. D’Agostino, O. Luongo, Model-independent reconstruction of $f(T)$ teleparallel cosmology, General Relativity and Gravitation, \textbf{49}, 141 (2017) \url{https://doi.org/10.1007/s10714-017-2304-x}
	%41
	%\bibitem{zhadyranova2024exploring} A. Zhadyranova, M. Koussour, S. Bekkhozhayev, V. Zhumabekova, J. Rayimbaev, Exploring late-time cosmic acceleration: A study of a linear $f(T)$ cosmological model using observational data, Physics of the Dark Universe, \textbf{45}, 101514 (2024) \url{https://doi.org/10.1016/j.dark.2024.101514}
	%42
	\bibitem{shekh2025cosmographical} S. H. Shekh, A. Pradhan, A. Dixit, S. N. Bayaskar, S. C. Darunde, Cosmographical analysis for $H(z)$ parametrization towards viscous $f(T)$ gravity, Modern Physics Letters A, \textbf{40}, 2450187 (2025) \url{https://doi.org/10.1142/S0217732324501876}
	%43
	\bibitem{duchaniya2024attractor} L. K. Duchaniya, K. Gandhi, B. Mishra, Attractor behavior of $f(T)$ modified gravity and the cosmic acceleration, Physics of the Dark Universe, \textbf{44}, 101461 (2024) \url{https://doi.org/10.1016/j.dark.2024.101461}
	%44
	\bibitem{maurya2023anisotropic} S. K. Maurya, A. Errehymy, M. Govender, G. Mustafa, N. Al-Harbi et al., Anisotropic compact stars in complexity formalism and isotropic stars made of anisotropic fluid under minimal geometric deformation (MGD) context in $f(T)$ gravity-theory, The European Physical Journal C, \textbf{83}, 348 (2023) \url{https://doi.org/10.1140/epjc/s10052-023-11507-w}
	%45
	\bibitem{maurya2022accelerating} D. C. Maurya, Accelerating scenarios of viscous fluid universe in modified $f(T)$ gravity, International Journal of Geometric Methods in Modern Physics, \textbf{19}, 2250144 (2022) \url{https://doi.org/10.1142/S0219887822501444}
	%46
	\bibitem{bamba2016bounce} K. Bamba, G. G. L. Nashed, W. El Hanafy, Sh. K. Ibraheem, Bounce inflation in $f(T)$ Cosmology: A unified inflaton-quintessence field, Physical Review D, \textbf{94}, 083513 (2016) \url{https://doi.org/10.1103/PhysRevD.94.083513}
	%48
	\bibitem{kavya2024can} N. S. Kavya, S. W. Mishra, P. K. Sahoo, V. Venkatesha, Can teleparallel $f(T)$ models play a bridge between early and late time Universe?, Monthly Notices of the Royal Astronomical Society, \textbf{532}, 3126--3133 (2024) \url{https://doi.org/10.1093/mnras/stae1723}
	%47
	\bibitem{bhar2024anisotropic} P. Bhar, F. Rahaman, S. Das, S. Aktar, A. Errehymy, Anisotropic quintessence compact star in $f(T)$ gravity with Tolman--Kuchowicz metric potentials, Communications in Theoretical Physics, \textbf{76}, 015401 (2024) \url{https://doi.org/10.1088/1572-9494/ad08ad}
	%49
	\bibitem{nunes2016new} R. C. Nunes, S. Pan, E. N. Saridakis, New observational constraints on $f(T)$ gravity from cosmic chronometers, Journal of Cosmology and Astroparticle Physics, \textbf{2016}, 011--011 (2016) \url{https://doi.org/10.1088/1475-7516/2016/08/011}
     %50
	\bibitem{chaudhary2024constraints} H. Chaudhary, U. Debnath, T. Roy, S. Maity, G. Mustafa et al., Constraints on the parameters of modified Chaplygin--Jacobi and modified Chaplygin--Abel gases in $f(T)$ gravity, International Journal of Geometric Methods in Modern Physics, \textbf{21}, 2450248 (2024) \url{https://doi.org/10.1142/S0219887824502487}
	%51
	\bibitem{duchaniya2022dynamical} L. K. Duchaniya, S. V. Lohakare, B. Mishra, S. K. Tripathy, Dynamical stability analysis of accelerating $f(T)$ gravity models, The European Physical Journal C, \textbf{82}, 448 (2022) \url{https://doi.org/10.1140/epjc/s10052-022-10406-w}
	%53
	\bibitem{chakraborty2023classical} M. Chakraborty, S. Chakraborty, The classical and quantum implications of the Raychaudhuri equation in $f(T)$-gravity, Classical and Quantum Gravity, \textbf{40}, 155010 (2023) \url{https://doi.org/10.1088/1361-6382/ace231}
	%54
	\bibitem{maurya2024role} S. K. Maurya, J. Kumar, S. Kiroriwal, Role of decoupling process on the configurations of compact stars induced by Thomas-Fermi dark matter with null complexity in $f(T)$ gravity, Journal of High Energy Astrophysics, \textbf{44}, 194--209 (2024) \url{https://doi.org/10.1016/j.jheap.2024.09.012}
	%55
	\bibitem{dixit2021probe} A. Dixit, A. Pradhan, D. C. Maurya, A probe of cosmological models in modified teleparallel gravity, International Journal of Geometric Methods in Modern Physics, \textbf{18}, 2150208 (2021) \url{https://doi.org/10.1142/S021988782150208X}
	%55
	\bibitem{patle2026revisiting} K. R. Patle, G. P. Singh, Revisiting $ f(T)$ Teleparallel Gravity with a Parametrized Hubble Parameter and Observational Constraints, arXiv preprint arXiv:2603.18971, (2026) \url{https://doi.org/10.48550/arXiv.2603.18971}
	%56
	\bibitem{das2023study} S. Das, A. Beesham, S. Chattopadhyay, Study of neutron star in $f(T)$ and $f(G)$ gravity framework with polytropic gas background, Annals of Physics, \textbf{458}, 169460 (2023) \url{https://doi.org/10.1016/j.aop.2023.169460}
	%57
	\bibitem{ren2022gaussian} X. Ren, S. F. Yan, Y. Zhao, Y. F. Cai, E. N. Saridakis et al., Gaussian processes and effective field theory of $f(T)$ gravity under the H 0 tension, The Astrophysical Journal, \textbf{932}, 131 (2023) \url{https://doi.org/10.3847/1538-4357/ac6ba5}
	%57
	\bibitem{riess2004type} A. G. Riess, L. G. Strolger, J. Tonry, S. Casertano, H. C. Ferguson, Type Ia supernova discoveries at z> 1 from the Hubble Space Telescope: Evidence for past deceleration and constraints on dark energy evolution, The Astrophysical Journal, \textbf{607}, 665--687 (2004) \url{https://doi.org/10.1086/383612}
	%57
	\bibitem{gong2006observational} Y. Gong, A. Wang, Observational constraints on the acceleration of the Universe, Physical Review D, \textbf{73}, 083506 (2006) \url{https://doi.org/10.1103/PhysRevD.73.083506}
	%57
	\bibitem{arora2024diagnostic} D. Arora, H. Chaudhary, S. K. J. Pacif, G. Mustafa, Diagnostic and comparative analysis of dark energy models with $q(z)$ parametrizations, The European Physical Journal Plus, \textbf{139}, 371 (2024) \url{https://doi.org/10.1140/epjp/s13360-024-05163-6}
	%57
	\bibitem{sanchez2011tracing} E. S{\'a}nchez, A. Carnero, J. Garc{\'\i}a-Bellido, E. Gaztanaga, F. De Simoni et al., Tracing the sound horizon scale with photometric redshift surveys, Monthly Notices of the Royal Astronomical Society, \textbf{411}, 277--288 (2011) \url{https://doi.org/10.1111/j.1365-2966.2010.17679.x}
	%57
	\bibitem{gadbail2022parametrization} G. N. Gadbail, S. Mandal, P. K. Sahoo, Parametrization of deceleration parameter in f(q) gravity, Physics, \textbf{4}, 1403--1412 (2022) \url{https://doi.org/10.3390/physics4040090}
	%57
	\bibitem{koussour2023modeling} M. Koussour, N. Myrzakulov, A. H. A. Alfedeel, F. Awad, M. Bennai, Modeling cosmic acceleration with a generalized varying deceleration parameter, Physics of the Dark Universe, \textbf{42}, 101339 (2023) \url{https://doi.org/10.1016/j.dark.2023.101339}
	%57
	\bibitem{shekh2024late} S. H. Shekh, K. S. Wankhade, S. N. Khan, A. Dixit, Late times $\Lambda$ CDM $f(T)$ gravity model with parameterized q(z), Modern Physics Letters A, \textbf{39}, 2450094 (2024) \url{https://doi.org/10.1142/S0217732324500949}
	%57
	\bibitem{al2016divergence} A. Al Mamon, S. Das, A divergence-free parametrization of deceleration parameter for scalar field dark energy, International Journal of Modern Physics D, \textbf{25}, 1650032 (2016) \url{https://doi.org/10.1142/S0218271816500322}
	%57
	\bibitem{xu2008constraints} L. Xu, H. Liu, Constraints to deceleration parameters by recent cosmic observations, Modern Physics Letters A, \textbf{23}, 1939--1948 (2008) \url{https://doi.org/10.1142/S0217732308025991}
	%16
	\bibitem{sofuouglu2026scalar} D. Sofuo{\u{g}}lu, B. Shukla, A. Beesham, Scalar Field Cosmology with Logarithmic Deceleration Parameter, in Proceedings of the 3rd International Online Conference on Universe, MDPI: Basel, Switzerland (2026)  
	%57
	\bibitem{samaddar2025reconstructing} A. Samaddar, S. S. Singh, Reconstructing cosmic expansion in $f(R, G)$ gravity using a log-periodic deceleration model, Physics of the Dark Universe, \textbf{50}, 102081 (2025) \url{https://doi.org/10.1016/j.dark.2025.102081}
	%58
	\bibitem{aldrovandi2012teleparallel} R. Aldrovandi, J. G. Pereira, Teleparallel gravity: an introduction, Springer Science \& Business Media, volume \textbf{173}, (2012) \url{https://doi.org/10.1007/978-94-007-5143-9}
	%59
	\bibitem{linder2010einstein} E. V. Linder, Einstein’s other gravity and the acceleration of the universe, Physical Review D, \textbf{81}, 127301 (2010) \url{https://doi.org/10.1103/PhysRevD.81.127301}
	%60
	\bibitem{maluf2013teleparallel} J. W. Maluf, The teleparallel equivalent of general relativity, Annalen der Physik, \textbf{525}, 339--357 (2013) 
	\url{https://doi.org/10.1002/andp.201200272}
	%61
	\bibitem{koussour2024exploring} M. Koussour, A. Altaibayeva, S. Bekov, F. Holmurodov, S. Muminov et al., Exploring cosmological evolution and constraints in $f(T)$ teleparallel gravity, Physics of the Dark Universe, \textbf{46}, 101664 (2024) \url{https://doi.org/10.1016/j.dark.2024.101664}
	%63
	\bibitem{banerjee2005acceleration} N. Banerjee, S. Das, Acceleration of the universe with a simple trigonometric potential, General Relativity and Gravitation, \textbf{37}, 1695–1703 (2005) \url{https://doi.org/10.1007/s10714-005-0152-6}
	%64
	\bibitem{cunha2008transition} J. Cunha, J. A. S. d. Lima, Transition redshift: new kinematic constraints from supernovae, Monthly Notices of the Royal Astronomical Society, \textbf{390}, 210–217 (2008) \url{https://doi.org/10.1111/j.1365-2966.2008.13640.x}
	%65
	\bibitem{escamilla2022dynamical} C. Escamilla-Rivera, A. N{\'a}jera, Dynamical dark energy models in the light of gravitational-wave transient catalogues, Journal of Cosmology and Astroparticle Physics, \textbf{2022}, 060 (2022) \url{https://doi.org/10.1088/1475-7516/2022/03/060}
	%68
	\bibitem{Zhao2006} W. Zhao, Y. Zhang, Quintom models with an equation of state crossing-1, Physical Review D, \textbf{73}, 123509 (2006) \url{https://doi.org/10.1103/PhysRevD.73.123509}
	%61
	\bibitem{foreman2013emcee} D. Foreman-Mackey, D. W. Hogg, D. Lang, J. Goodman, emcee: the MCMC hammer, Publications of the Astronomical Society of the Pacific, \textbf{125}, 306 (2013) \url{https://doi.org/10.1086/670067}
	%62
	\bibitem{simon2005constraints} J. Simon, L. Verde, R. Jimenez, Constraints on the redshift dependence of the dark energy potential, Physical Review D, \textbf{71}, 123001 (2005) \url{https://doi.org/10.1103/PhysRevD.71.123001}
	%63
	\bibitem{sharov2018predictions} G. S. Sharov, V. O. Vasiliev, How predictions of cosmological models depend on Hubble parameter data sets, arXiv preprint arXiv:1807.07323 (2018) \url{https://doi.org/10.26456/mmg/2018-611}
	%64
	\bibitem{stern2010cosmic} D. Stern, R. Jimenez, L. Verde, M. Kamionkowski, S. A. Stanford, Cosmic chronometers: constraining the equation of state of dark energy. I: $H(z)$ measurements, Journal of Cosmology and Astroparticle Physics, \textbf{2010}, 008 (2010) \url{https://doi.org/10.1088/1475-7516/2010/02/008}
	%65
	\bibitem{moresco2015raising} M. Moresco, Raising the bar: new constraints on the Hubble parameter with cosmic chronometers at $z~ 2$, Monthly Notices of the Royal Astronomical Society: Letters, \textbf{450}, L16--L20 (2015) \url{https://doi.org/10.1093/mnrasl/slv037}
	%66
	\bibitem{jimenez2002constraining} R. Jimenez, A. Loeb, Constraining cosmological parameters based on relative galaxy ages, The Astrophysical
	Journal, \textbf{573}, 37 (2002) \url{https://doi.org/10.1086/340549}
	%68
	\bibitem{mandal2023cosmic} S. Mandal, A. Singh, R. Chaubey, Cosmic evolution of holographic dark energy in $f(Q, T)$ gravity, International Journal of Geometric Methods in Modern Physics, \textbf{20}, 2350084 (2023) \url{https://doi.org/10.1142/S0219887823500846}
	%68
	\bibitem{garg2025cosmological} R. Garg, T. Chowdhury, G. P. Singh, F. Rahaman, Cosmological model with Gong-Zong parametrization in $f(R, Lm)$ gravity, Physics of the Dark Universe, \textbf{49}, 102025 (2025) \url{https://doi.org/10.1016/j.dark.2025.102025}
	%69
	\bibitem{scolnic2018complete} D. M. Scolnic, D. O. Jones, A. Rest, Y. C. Pan, R. Chornock et al., The complete light-curve sample of spectroscopically confirmed SNe Ia from Pan-STARRS1 and cosmological constraints from the combined Pantheon sample, The Astrophysical Journal, \textbf{859}, 101 (2018) \url{https://doi.org/10.3847/1538-4357/aab9bb}
	%70
	\bibitem{riess1999bvri} A. G. Riess, R. P. Kirshner, B. P. Schmidt, S. Jha, P. Challis et al., BVRI light curves for 22 type Ia supernovae, The Astronomical Journal, \textbf{117}, 707 (1999) \url{https://doi.org/10.1086/300738}
	%71
	\bibitem{hicken2009improved} M. Hicken, W. M. Wood-Vasey, S. Blondin, P. Challis, S. Jha et al., Improved dark energy constraints from ~100 new CfA supernova type Ia light curves, The Astrophysical Journal, \textbf{700}, 1097 (2009) \url{https://doi.org/10.1088/0004-637X/700/2/1097}
	%72
	\bibitem{sako2018data} M. Sako, B. Bassett, A. C. Becker, P. J. Brown, H. Campbell et al., The data release of the Sloan Digital Sky Survey-II supernova survey, Publications of the Astronomical Society of the Pacific, \textbf{130}, 064002 (2018) \url{https://doi.org/10.1088/1538-3873/aab4e0}
	%73
	\bibitem{guy2010supernova} J. Guy, M. Sullivan, A. Conley, N. Regnault, P. Astier, et al., The Supernova Legacy Survey 3-year sample: Type Ia supernovae photometric distances and cosmological constraints, Astronomy $\&$ Astrophysics, \textbf{523},  (2010) \url{https://doi.org/10.1051/0004-6361/201014468}
	%74
	\bibitem{contreras2010carnegie} C. Contreras, M. Hamuy, M. M. Phillips, G. Folatelli, N. B. Suntzeff, et al., The Carnegie Supernova Project: first photometry data release of low-redshift type Ia supernovae, The Astronomical Journal, \textbf{139}, 519 (2010) \url{https://doi.org/10.1088/0004-6256/139/2/519}
	%75
	\bibitem{odintsov2018cosmological} S. D. Odintsov, V. Oikonomou, A. Timoshkin, E. N. Saridakis, R. Myrzakulov, Cosmological fluids with logarithmic equation of state, Annals of Physics, \textbf{398}, 238--253 (2018) \url{https://doi.org/10.1016/j.aop.2018.09.015}
	%85
	\bibitem{ellis2012relativistic} G. F. R. Ellis, R. Maartens, M. A. H. MacCallum, Relativistic cosmology, Cambridge University Press, (2012) \url{http://dx.doi.org/10.1017/CBO9781139014403}	
	%76
	\bibitem{asvesta2022observational} K. Asvesta, L. Kazantzidis, L. Perivolaropoulos, C. G. Tsagas, Observational constraints on the deceleration parameter in a tilted universe, Monthly Notices of the Royal Astronomical Society, \textbf{513}, 2394--2406 (2022) \url{https://doi.org/10.1093/mnras/stac922}	
	%81
	\bibitem{visser1997energy} M. Visser, Energy conditions in the epoch of galaxy formation, Science, \textbf{276}, 88--90 (1997) \url{https://doi.org/10.1126/science.276.5309.88}
	%82
	\bibitem{lalke2024cosmic} A. R. Lalke, G. P. Singh, A. Singh, Cosmic dynamics with late-time constraints on the parametric deceleration parameter model, European Physical Journal Plus, \textbf{139}, 288 (2024) \url{https://doi.org/10.1140/epjp/s13360-024-05091-5}
	%83
	\bibitem{singh2022lagrangian} A. Singh, R. Raushan, R. Chaubey, S. Mandal, K. C. Mishra, Lagrangian formulation and implications of barotropic fluid cosmologies, International Journal of Geometric Methods in Modern Physics, \textbf{19}, 2250107 (2022) \url{https://doi.org/10.1142/S0219887822501079}
	%84
	\bibitem{mishra2025cosmological} S. S. Mishra, P. K. Sahoo, Cosmological aspects in the constrained $f(T, T)$ theory using Raychaudhuri equations, Physics of the Dark Universe, \textbf{48}, 101887 (2025) \url{https://doi.org/10.1016/j.dark.2025.101887}
	%66
	%\bibitem{myrzakulov2023quintessence} N. Myrzakulov, M. Koussour, A. Mussatayeva, Quintessence-like features in the late-time cosmological evolution of $f(Q)$ symmetric teleparallel gravity, Chinese Journal of Physics, \textbf{85}, 345--358 (2023) \url{https://doi.org/10.1016/j.cjph.2023.07.003}
	%84
	\bibitem{Sahni2003} V. Sahni, T. D. Saini, A. A. Starobinsky, U. Alam, Statefinder-a new geometrical diagnostic of dark energy, Journal of Experimental and theoretical Physics Letters, \textbf{77}, 201–206 (2003) \url{https://doi.org/10.1134/1.1574831}
	%84
	\bibitem{fei2013statefinder} Y. Fei, Z. Jing-Fei, Statefinder diagnosis for the extended holographic Ricci dark energy model without and with interaction, Communications in Theoretical Physics, \textbf{59}, 243--248 (2013) \url{https://doi.org/10.1088/0253-6102/59/2/17}
	%77
	\bibitem{sahni2008two} V. Sahni, A. Shafieloo, A. A. Starobinsky, Two new diagnostics of dark energy, Physical Review D, \textbf{78}, 103502 (2008) \url{https://doi.org/10.1103/PhysRevD.78.103502}
	 %89
	\bibitem{bamba2012reconstruction} K. Bamba, R. Myrzakulov, S. Nojiri, S. D. Odintsov, Reconstruction of $f(T)$ gravity: Rip cosmology, finite-time future singularities,<? format?> and thermodynamics, Physical Review D, \textbf{85}, 104036 (2012) \url{https://doi.org/10.1103/PhysRevD.85.104036}
	%89
	\bibitem{tong2009cosmic} M. L. Tong, Y. Zhang, Cosmic age, statefinder and Om diagnostics in the decaying vacuum cosmology, Physical Review D, \textbf{80}, 023503 (2009) \url{https://doi.org/10.1103/PhysRevD.80.023503}
	%89
	\bibitem{tatum2025extracting} E. T. Tatum, E. G. Haug, Extracting a Cosmic Age of 14.6 Billion Years from All 580 Supernova Redshifts in the Union2 Database, Journal of Modern Physics, \textbf{16}, 507--517 (2025) \url{https 10.4236/jmp.2025.164026}
	%89
	\bibitem{Valcin2020} D. Valcin, J. L. Bernal, R. Jimenez, L. Verde, B. D. Wandelt, Inferring the age of the universe with globular clusters, Journal of Cosmology and Astroparticle Physics, \textbf{12}, 002-002 (2020) \url{https://doi.org/10.1088/1475-7516/2020/12/002}
	
	
%%%%%%%%%%%%%%%%%%%%%%%%%%%%%%%%%%%%%%%%%%%%%%%%%%%%%%%%%%%%%%%%%%%%%%%%%%%%%	
	 	 
\end{thebibliography}
\end{document}